\documentclass{aastex631}

\usepackage{amsmath}

\begin{document}

\title{Identification of Nonlinear Damping of Transverse Loop Oscillations by KHI-induced Turbulence}

\correspondingauthor{Sihui Zhong}
\email{s.zhong3@exeter.ac.uk}

\author[0000-0002-5606-0411]{Sihui Zhong}
\affiliation{Department of Mathematics and Statistics, University of Exeter,
Exeter, EX4 4QF, UK}

\author[0000-0002-0851-5362]{Andrew Hillier}
\affiliation{Department of Mathematics and Statistics, University of Exeter,
Exeter, EX4 4QF, UK}

\author[0000-0002-7008-7661]{I\~nigo Arregui}
\affiliation{Instituto de Astrof\'{\i}sica de Canarias, V\'ia L\'actea S/N, E-38205 Laguna, Tenerife, Spain}
\affiliation{Departmento de Astrof\'{\i}sica, Universidad de La Laguna, E-38206 Laguna, Tenerife, Spain}

\begin{abstract}
Kink oscillations in coronal loops have been extensively studied for their potential contributions to coronal heating and their role in plasma diagnostics through coronal seismology. A key focus is the strong damping of large-amplitude kink oscillations, which observational evidence suggests is nonlinear. However, directly identifying the nonlinearity is a challenge.
This work presents an analytic formula describing nonlinear standing kink oscillations dissipated by turbulence, characterised by a time-varying damping rate and period drift. We investigate how the damping behaviour depends on the driving amplitude and loop properties, showing that the initial damping time $\tau$ is inversely proportional to the velocity disturbance over the loop radius, $V_i/R$.
Using MCMC fitting with Bayesian inference, the nonlinear function better fits an observed decaying kink oscillation than traditional linear models, including exponential damping, suggesting its nonlinear nature. 
By applying a Bayesian model comparison, we establish regimes in which nonlinear and linear resonant absorption mechanisms dominate based on the relationship between the damping rate $\tau/P$ and $V_i/R$. 
Additionally, analysis of two specific events reveals that while one favours the nonlinear model, the other is better explained by the linear model.
Our results suggest that this analytical approximation of nonlinear damping due to turbulence provides a valid and reliable description of large-amplitude decaying kink oscillations in coronal loops. 
%Using the nonlinear damping model and the linear resonant absorption model to explain the relationship of damping rate $\tau/P$ and $V_i/R$, we identify the regimes dominated by nonlinearity and linearity using Bayesian model comparison.

\end{abstract}

%% The AAS Journals now uses Unified Astronomy Thesaurus concepts: https://astrothesaurus.org
%% You will be asked to select these concepts during the submission process


\section{Introduction} \label{sec:intro}
%First, a bit of history of transverse (kink) oscillations.
Since the discovery of transverse coronal loop oscillations \citep{1999Sci...285..862N,1999ApJ...520..880A} by the \textit{Transition Region and Coronal Explorer} (TRACE), this phenomenon has been routinely detected and studied \citep[e.g.][and references therein]{2024RvMPP...8...19N}. This wave phenomenon typically occurs after impulsive eruptions such as flares and undergoes rapid decay within a few cycles. They are characterised by the periodic transverse displacement of the loop axis and are thus interpreted as kink modes (m=1) in a magnetised plasma cylinder within the magnetohydrodynamics (MHD) framework. A statistical study of decaying kink oscillations using a catalogue of events \citep{2019ApJS..241...31N} (hereafter referred to as the 2019 Catalogue) established a linear scaling between the oscillation period and the loop length, indicating the standing-wave nature of kink oscillations. 
Since their initial discovery, kink oscillations have become a key topic in the solar physics community due to their potential role in coronal wave heating mechanisms \citep{2020SSRv..216..140V} and coronal seismology \citep{2021SSRv..217...73N,2024RvMPP...8...19N}, where plasma parameters are inferred from the observed MHD wave properties.

In the context of coronal heating, intensive studies on decaying kink oscillations have focused on explaining their strong damping. In the linear MHD regime, kink oscillations in a loop with a smooth transition in Alfv\'en speed between the loop and the external medium are damped by resonant absorption \citep{2002A&A...394L..39G,2002ApJ...577..475R}.
This mechanism involves the resonance between the global kink mode and local Alfv\'en mode, which results in the transfer of energy from the global oscillation to small-scale azimuthal motions in a resonance layer around the location where the Alfv\'en frequency matches the kink frequency.  
In the framework of linear resonant absorption, the time variation of oscillation amplitude is described by a kind of Gaussian profile for a small time (i.e. at the beginning of oscillation) and by the exponential function for larger time \citep{2012A&A...539A..37P,2013A&A...555A..27R}. This also applies to the spatial damping of propagating kink waves \citep{2013A&A...551A..39H}. The transition between Gaussian and exponential damping depends on the thickness of the transition layer. In any case, the damping profile is independent of the oscillation amplitude in the linear regime.

Beyond the linear regime, nonlinear damping has been considered, supported by the following observational evidence.
(1) Empirically, 94\% of the detected 252 events exhibit oscillation amplitudes greater than 1~Mm \citep[e.g.,][]{2019ApJS..241...31N} which is the typical loop radius. This challenges the validity of the small-amplitude assumption of linear MHD wave theory. In the large amplitude regime, wave dynamics -- such as the generation of higher-order modes and instabilities, as well as consequent enhancement of damping -- becomes amplitude-dependent \citep[e.g.][]{2010PhPl...17h2108R,2025A&A...693A.201S}, a key aspect that classic linear wave theory fails to account for. %Indeed, large-amplitude oscillations evolve nonlinearly (Ruderman et al.2010, 2014,2017?) \citep{2016A&A...595A..81M}. % Large amplitude will introduce higher-order modes, interacting with the kink mode, hence causing intensity modification, additional decay [cite] etc.
(2) The oscillatory pattern is not purely harmonic (see Figure 2 in \citealt{2019A&A...632A..64D}), it may result from the superposition of multiple harmonics or the effects of nonlinearities \citep{1997SoPh..175...93N}. 
(3) The damping of kink oscillations depends on amplitude \citep{2016A&A...590L...5G}. When the damping ratio (also called the quality factor, defined as the ratio of damping time to oscillation period, $Q=\tau/P$) for a large number of loop oscillation events in 2019
Catalogue is plotted against their oscillation amplitude ($\xi$), the data are scattered forming a cloud with a
triangular shape. Larger amplitudes correspond, in general, to smaller damping ratio values and vice versa. Fitting the triangle-shaped cloud boundary gives an amplitude dependence of the quality factor of $Q = \xi^{-0.68}$ \citep{2019ApJS..241...31N}.
%Specifically, the quality factor, defined as the ratio of damping time to oscillation period, $Q=\tau/P$, scales with amplitude $\xi$ as $Q = \xi^{-0.68}$, as revealed by \citet{2019ApJS..241...31N}.
(4) The damping profile deviates from an exponential form. \citet{2023MNRAS.525.5033Z} re-evaluated the damping patterns of decaying kink oscillations and proposed a super-exponential model as a guess of nonlinearity. Their analysis showed that 7 out of 10 events are better described by a super-exponential function.

Theoretically, nonlinearity in impulsively driven kink oscillation leads to frequency drift \citep{2014SoPh..289.1999R}, the ponderomotive force and consequently mass accumulation at the loop apex \citep{2004ApJ...610..523T}. Moreover, it excites sausage and fluting modes ($m\geq 2$)\citep[e.g.][]{2009ApJ...694..502O,2017ApJ...836..219A, 2017SoPh..292..111R,2018ApJ...853...35T}. The damping of the oscillation can be enhanced by resonance between the kink mode and shorter-wavelength fluting modes, which makes energy go from the former to the latter. However, $m\geq2$ mode resonance is suppressed when gravity stratification in the loop is considered. An analytical estimate of the damping rate due to nonlinear cascade to higher order modes was derived by \citet{2021ApJ...910...58V} using Els\"asser variables. Their model predicts that the quality factor is inversely proportional to the ratio of displacement amplitude to loop radius, and is also dependent on the density contrast. Their predicted scaling with $Q\propto \xi^{-1}$ seems to qualitatively match the observations of \cite{2019ApJS..241...31N} well.
When driving velocity is sufficiently high, Kelvin-Helmholtz instability (KHI) \citep{1983A&A...117..220H,2008ApJ...687L.115T} develops at the loop boundaries due to the shear between the high-density oscillating loop and the low-density stationary surroundings. This manifests as small-scale vortices rolling around the loop, eventually evolving into turbulence \citep[e.g.][]{2016ApJ...830L..22A,2024ApJ...966...68H}. As a result, wave energy from the global kink mode cascades to smaller scales, accelerating oscillation decay. 
A parametric numerical study on KHI in standing kink oscillations by \citet{2016A&A...595A..81M} demonstrates that higher velocity amplitudes lead to faster damping.

So far, there has been no direct observational evidence of nonlinearity in transverse coronal loop oscillations.
Theoretically predicted signatures, such as short-wavelength fluting modes or small-scale KHI vortices, remain below the spatial resolution of current solar instrumentation. Furthermore, since the heat generated by KHI-induced turbulence is minimal \citep{2016A&A...595A..81M,2017A&A...607A..77H,2024ApJ...966...68H}, oscillating loops do not exhibit significant enhancement in brightness.
The lack of directly detectable nonlinear features in kink oscillations poses a significant challenge in validating nonlinear wave theories. Misinterpretation of observed waves directly impacts the accuracy of plasma diagnostics, potentially leading to misestimates of coronal properties. Identifying signatures of nonlinear damping in kink oscillations is therefore crucial -- not only for improving the reliability of seismological techniques but also for assessing its role in coronal wave heating. 
One notable attempt by \citet{2021ApJ...915L..25A} used Bayesian model comparison to evaluate the nonlinear damping model by \citet{2021ApJ...910...58V} in front of the exponential resonant damping model in a sample of 101 observed loop oscillations. Their study found that the marginal likelihood for the nonlinear damping model was larger than that for resonant damping for the majority of cases, 85, with positive evidence in favour of the nonlinear model in 60 of them. %determined by the resonant absorption. Their study found that 60 out of 101 observed oscillations prefer nonlinear damping through uniturbulence model with conclusive evidence.
However, the damping time used as input data was measured assuming exponential decay based on linear theory, the discrepancy between linear and nonlinear damping time remains to be determined.

Using 3D MHD simulations and analytical calculations, \citet{2024ApJ...966...68H} investigated the evolution of KHI-induced turbulence on a flux tube, where a nonlinear fundamental kink mode was excited. They made accurate predictions regarding the damping of the oscillation and turbulent heating as consequences of the KHI dynamics. Their results showed damping profiles distinct from those of linear theory, which is commonly used to compare theoretical models with observations. Building on these results, our aim here is to identify the signatures of nonlinear damping in observed transverse loop oscillations. Because the direct comparison between 3D MHD simulations and observations is challenging, we first provide an analytical expression for the time evolution of the nonlinear oscillation amplitude, illustrate its dependence on the relevant parameters, and highlight its observational features (Section~\ref{sec:model}). Our data analysis methods are described in Section~\ref{sec:methods}. The model is then validated observationally through curve fitting with the time series and Bayesian model comparison of the quality factor (Section~\ref{sec:validation}). In our study, we base our judgment on the evidence in favour of any of the models under consideration on the quantification of Bayesian evidence. Finally, we discuss the results in Section~\ref{sec:dis}.

%Here, we investigate the nonlinear damping of standing kink oscillation in a coronal loop due to KHI-induced turbulence. This study builds on \citet{2024ApJ...966...68H}, which studies the KHI dynamics using 3D MHD simulations and analytical calculations.
%In this work, we provide a quantitative description of the nonlinear oscillation time series, distinguishing it from the linear damping profile. In addition, we determine the damping time during the first oscillation cycle. The model is validated observationally through curve fitting with the time series and Bayesian model comparison of the quality factor.

\section{Nonlinear damping by KHI-induced turbulence}\label{sec:model}

The physics of the mixing process between two magnetic layers with different densities has been formulated (see \citealt{2019ApJ...885..101H}), and its effects on an oscillating tube have been investigated in \citet{2024ApJ...966...68H}.
An impulsive transverse perturbation is applied to a straight magnetic tube with higher density, embedded in a rarefied background, to generate a fundamental kink oscillation. The transverse density profile of the tube is set to be very sharp, preventing resonant absorption during the initial phase of the oscillation. Once the transverse oscillation begins, the shear flow between the loop and its surroundings leads to material mixing in two layers, forming vortices and, consequently, a mixing/turbulence layer around the tube. %This process is known as the Kelvin-Helmholtz instability (KHI).
More importantly, as shown in \citet{2024ApJ...966...68H}, this mixing extracts momentum from the tube core and transfers it to the mixing layer. The turbulence layer grows self-similarly until the entire tube becomes turbulent \citep{2019FrP.....7...85A}. This density mixing and momentum exchange damp the oscillation during the turbulence development stage. 

\begin{figure}[h!]
	\centering
	\includegraphics[width=0.6\linewidth]{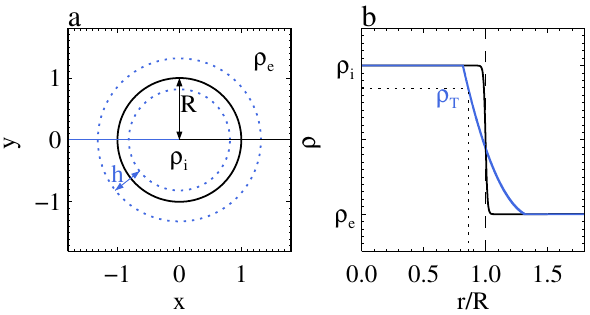}
	\caption{
        Sketch of the loop cross-section (panel a) and transverse density profile (panel b) at the initial moment (black) and at the KHI developing stage (blue). In panel a, the density inside/outside the cross-section of the loop is $\rho_i$/$\rho_e$. The loop radius is $R$. The black circle denotes the initial density contour of $\rho_i$. The blue circles outline the mixing layer developed in later times, its height is $h$. The transverse density profile across the loop is shown in panel b, with the black representing the initial distribution and the blue one at a certain time when $h$ reaches half of the loop radius. The upper density limit, $\rho_T$, is used to confine the area of the mixing layer in our formula.
	}
	\label{fig:sktech}
\end{figure}

\subsection{Analytic formula}
\label{sec:formula}

%Based on \citet{2024ApJ...966...68H}, we derive the analytic formula for the time-evolving amplitude of a transversely oscillating straight magnetic loop damped by KHI-induced turbulence. 
\citet{2024ApJ...966...68H} showed that two distinct stages characterise the evolution of KHI-induced turbulence in a transversely oscillating tube. In the first stage, the turbulent layer grows in a self-similar manner, scaling linearly with time ($\propto t$), a property that can be used to predict oscillation damping. In the second stage, the turbulent energy decays as $t^{-2}$, providing a way to estimate the heating rate. In this study, we focus on the first stage in the context of a straight magnetic loop, review the derivation of the loop’s velocity and further derive an analytic formula for the time evolution of a transversely oscillating loop damped by KHI-induced turbulence.

First, let us introduce parameters that describe the loop system. For the loop properties, we consider the loop radius $R$, internal/external density $\rho_{i}$/$\rho_{e}$, and the global kink frequency of the loop $\omega_k$ calculated using linear wave theory. 
The magnetic field is uniform and directed along the loop axis. For the external perturbation to trigger the transverse oscillation, we take the initial velocity perturbation $V_{i}$. We confine the mixing layer by $\rho_{e}$ and the upper density limit/threshold $\rho_T$ with a height of $h$, see Fig.~\ref{fig:sktech}. $h$ reaches a maximum value $H=\pi R/4$ when the whole tube becomes turbulent. This is calculated by approximating, purely for the turbulent transport terms, the circular loop cross-section as a rectangular one with the same mass and momentum. This is just a mathematical tool to allow the mixing layer model to be combined with an oscillating tube and has no impact on the physics of an oscillating tube.
The spatially-averaged density profile of the mixing layer (see Fig.~\ref{fig:sktech}b) can be estimated by assuming it is a third-order polynomial and applying three conditions (see details in Section 2.3.1 of \citealt{2019ApJ...885..101H}). Once the density profile and $\rho_T$ are determined, the Alfv\'en frequency in the mixing layer, $\omega_A$, is approximated as the mean of the inverse square root of the density, multiplied by the wavenumber and magnetic field. Here we define $\omega_{\rm S} = (\omega_{\rm k} + \omega_{\rm A})/2, \omega_{\rm D} = (\omega_{\rm k} - \omega_{\rm A})/2$ as the average oscillation frequency and the difference frequency between the oscillations of the loop core and mixing layer to characterise their interference pattern.  
Lastly, $C_1$ is a constant to describe the mixing physics in hydrodynamics. As predicted by \citet{2019ApJ...885..101H,2023MNRAS.520.1738H}, the height of the mixing layer grows in a self-similar manner in the form of
\begin{equation}\label{eq:height}
    h(t)=\frac{C_1}{2}\frac{(\rho_i \rho_e)^{1/4}}{\sqrt{\rho_i} + \sqrt{\rho_e}}\Delta V t, 
\end{equation}
\noindent where $\Delta V$ is the magnitude of the shear flow and $(\rho_i \rho_e)^{1/4}\Delta V/[2(\sqrt{\rho_i} + \sqrt{\rho_e})]$ is the root-mean-squared velocity of the turbulent motions. During the mixing process, $\Delta V$ is assumed to be constant in time. %However, $\Delta V$ is varying in time as observed in simulation, so we take a time-averaged value as representative here.
Given that the mixing layer grows $\propto t$, the mass change rate above a density threshold of the mixing layer in a unit of width, $\Delta m_{\rho>\rho_T}$, can be calculated for a given density profile \citep{2019ApJ...885..101H}. %with negative means mass loss and positive indicates gain.

The time evolution of the mass, momentum and energy in the oscillating loop, along with all necessary assumptions and their justification, are fully detailed in Sections~3.1-3.2 and Section~4 of \citet{2024ApJ...966...68H}. For conciseness, we do not repeat all of them here.
%\sout{Based on the theory, the mass and momentum of the tube and hence the amplitude of the oscillation can be derived with the above parameters, see more details in Sections~3.1- 3.2 in \citet{2024ApJ...966...68H}.} 
As KHI develops in a self-similar manner, the time evolution of mass change depends linearly on the mass change rate, $\Delta m_{\rho>\rho_T}$, and the growth rate of the mixing layer height (Eq.~\ref{eq:height}). Thus, the mass in the tube core above a given density threshold $\rho_T$ is given by 
\begin{equation}
    m_{\rho>\rho_T}(t) = 4R H \rho_i\left(1 +\frac{\Delta m_{\rho>\rho_T} C_1}{2H \rho_i}\frac{(\rho_i \rho_e)^{1/4}}{\sqrt{\rho_i} + \sqrt{\rho_e}}\Delta V t  \right), %\frac{C_2}{\sqrt{2}}
\end{equation}
\noindent where $4RH\rho_i$ is the initial mass of the tube. %Usually, $C_1$ is multiplied by $\sqrt{2}$ to take into account that the magnitude of the shear flow is not uniform in space.
The tube core is a forced linear oscillator with natural frequency $\omega_{\rm k}$, driven by the initial momentum {$4RH\rho_i V_i$}, and damped through the momentum extraction with the rate of $dh/dt$. Note that to calculate momentum transport we model the mixing layer as being in the zero-momentum frame and the proportion of this layer that falls on the high density side (i.e. the tube core) is given by $ \sqrt{ \rho_e} / (\sqrt{\rho_i} + \sqrt{\rho_e}) $ \citep{2019ApJ...885..101H}. Therefore, the momentum in the tube core ${p}_{\rm core}$ is given by
\begin{equation}
{p}_{\rm core}(t) =  4R H \rho_i V_i \cos(\omega_{\rm k}t)\left(1 -\frac{C_1}{2H}\frac{(\rho_i \rho_e)^{1/4}}{\sqrt{\rho_i} + \sqrt{\rho_e}}\frac{\sqrt{\rho_e}}{\sqrt{\rho_i} + \sqrt{\rho_e}}\Delta V t  \right). \label{eq:M_core}\\
\end{equation}
Note that a factor of $\sqrt{\rho_e}/(\sqrt{\rho_i}+\sqrt{\rho_e})$ is missing in Eq.(18) in \citet{2024ApJ...966...68H}. This factor appears in later equations, so the function in the paper is correct.
The mixing layer is treated as a driven oscillator %with local Alfv\'en frequency $\omega_{\rm A}$, taking oscillatory forcing from the loop core with $\omega_{\rm k}$, which grows at the same rate of $dh/dt$. 
with frequency of $\omega_{\rm S}$, modulated by an envelope varying at $\omega_{\rm D}t$. In addition, the rate of momentum forcing in the layer is proportional to $dh/dt$. Therefore, the momentum in the mixing layer ${p}_{\rm L}$ is given by
\begin{align}
{p}_{\rm L}(t)\approx 2\overline{p}C_1\frac{(\rho_i \rho_e)^{1/4}}{\sqrt{\rho_i} + \sqrt{\rho_e}}\frac{\Delta V R}{\omega_{\rm D}}\cos(\omega_{\rm S} t)\sin(\omega_{\rm D} t), \label{eq:M_L}
\end{align}
\noindent with
\begin{equation}
    \overline{p}=\left(V_i-\frac{\Delta V\sqrt{\rho_e}}{\sqrt{\rho_i}+\sqrt{\rho_e}}\right)\times \int_{y'(\rho_T)}^{y'(\rho_i)} \rho dy',
\end{equation}
\noindent where $\int_{y'(\rho_T)}^{y'(\rho_i)} \rho dy'$ is the mass per unit width between the core and the mixing layer, defined as $\Xi$, and $\overline{p}$ is the momentum injection per unit width. Eq.~\ref{eq:M_L} is satisfied when $\omega_{\rm D}\ll \omega_{\rm S}$, that is, $\omega_{\rm k}< \omega_{\rm A}$. 

Let us set 
\begin{align}
E = & 4RH\rho_{i}, \\
A = & E V_{i}, \\
B = & \frac{C_1}{2H}\frac{\sqrt{\rho_e}}{\sqrt{\rho_i} + \sqrt{\rho_e}}\frac{(\rho_i \rho_e)^{1/4}}{\sqrt{\rho_i} + \sqrt{\rho_e}}\Delta V,\\
D = & 2\overline{p}C_1\frac{(\rho_i \rho_e)^{1/4}}{\sqrt{\rho_i} + \sqrt{\rho_e}}\frac{\Delta V R}{\omega_{\rm D}},\\
G = & \frac{\Delta m_{\rho>\rho_T}C_1}{2H \rho_i}\frac{(\rho_i \rho_e)^{1/4}}{\sqrt{\rho_i} + \sqrt{\rho_e}}\Delta V. 
\end{align}
Then the centre-of-mass velocity, which is defined as the total momentum of the loop system divided by the mass, can be written as:
\begin{align}\label{eq:velocity_amp}
    V_{\mathrm{CoM}} = & \frac{{p}_{\rm core}(t)+{p}_{\rm L}(t)}{m_{\rho>\rho_T}(t)} \nonumber\\
            = & \frac{A \cos(\omega_{\rm k}t)(1-Bt) + D\cos(\omega_{\rm S} t)\sin(\omega_{\rm D} t)}{E(1+Gt)}.
\end{align}
To obtain the amplitude of the oscillating loop, we integrate $V_{\mathrm{CoM}}$ over time
\begin{align}
    \xi_{\rho>\rho_T}(t) = & \int_0^t \frac{A \cos(\omega_{\rm k}t)(1-Bt) + D\cos(\omega_{\rm S} t)\sin(\omega_{\rm D} t)}{E(1+Gt)}dt \nonumber\\
    = & \frac{A}{E G^2}\left( -\frac{BG}{\omega_{\rm k}}\sin(\omega_{\rm k}t) + (B+G)\left(\cos\left(\frac{\omega_{\rm k}}{G} \right){\rm Ci}\left[\omega_{\rm k}\left( t+\frac{1}{G}\right) \right] +\sin\left(\frac{\omega_{\rm k}}{G} \right){\rm Si}\left[\omega_{\rm k}\left( t+\frac{1}{G}\right) \right]\right) \right) \nonumber\\
    & +\frac{D}{2E G} \left(-\cos\left(\frac{\omega_{\rm k}}{G} \right){\rm Si}\left[\omega_{\rm k}\left( t+\frac{1}{G}\right) \right]+\sin\left(\frac{\omega_{\rm k}}{G} \right){\rm Ci}\left[\omega_{\rm k}\left( t+\frac{1}{G}\right) \right]\right. \nonumber\\
    & \left. +\cos\left(\frac{\omega_{\rm A}}{G} \right){\rm Si}\left[\omega_{\rm A}\left( t+\frac{1}{G}\right) \right]-\sin\left(\frac{\omega_{\rm A}}{G} \right){\rm Ci}\left[\omega_{\rm A}\left( t+\frac{1}{G}\right) \right]\right) + \mathrm{Constant}. \label{eq:nl_amplitude}
\end{align}
This function involves the sine integral function ${\rm Si}(x)=\int^{x}_{0} \frac{\sin(t)}{t}dt$ and cosine integral function ${\rm Ci}(x)=-\int^{\infty}_{x} \frac{\sin(t)}{t}dt$, see examples in Appendix~\ref{Appendix:si(x)}. The function is not defined at $t_G=-1/G$, that is, at zero mass above a given $\rho_T$. Also, according to Eq.~\ref{eq:M_core}, the momentum of the loop core becomes zero at $t_B=1/B$, this is also the time when the whole tube becomes turbulent. Both time scales are inversely proportional to $\Delta V/R$. With the typical density threshold, e.g., $\rho_T> 0.55\rho_i, t_B < t_G$. Eq.~\ref{eq:nl_amplitude} gives the analytic expressions for the time evolution of the oscillation amplitude in our nonlinear model and is our first main result. For simplicity, we divide all density-related parameters by $\rho_e$, yielding the same expressions as in Eqs.~\ref{eq:velocity_amp}–\ref{eq:nl_amplitude}. This highlights that the density ratio $\rho_i/\rho_e$ governs the oscillation behaviour. We adopt normalised terms in the following analysis. %[Ruderman+2010,2014 the eigenfunctions.]

\begin{figure}[!ht]
	\centering
	\includegraphics[width=0.9\linewidth]{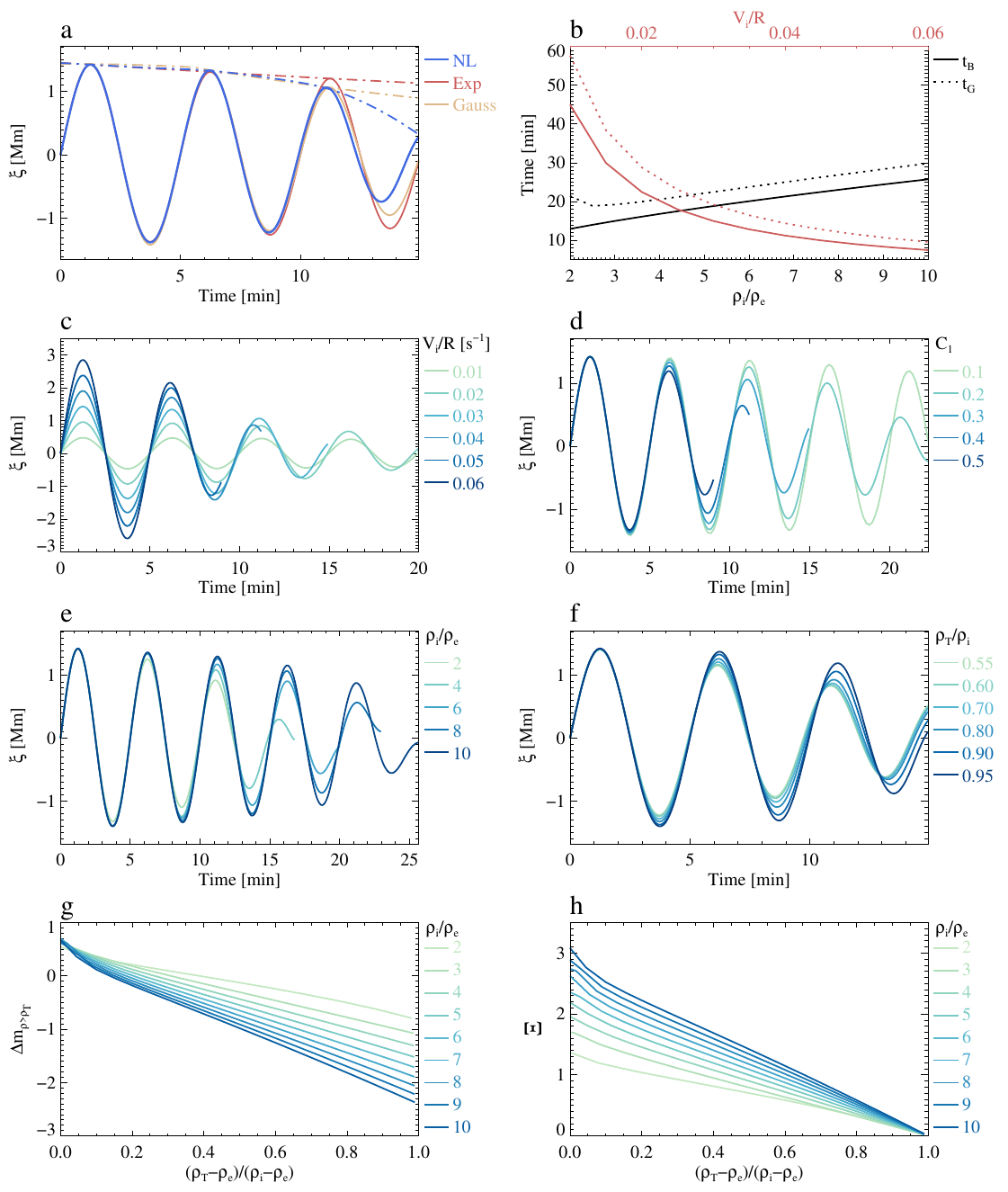}
	\caption{
        Characteristics of nonlinear transverse oscillations of a 200 Mm-long loop described by Eq.~\ref{eq:nl_amplitude}. The blue curve in panel a is constructed using the following values: $R=1$ Mm, $\omega_{\rm k}=0.02$ Hz, $\omega_{\rm A}=0.023$ Hz, $\rho_i/\rho_e=3$, $C_1=0.3$, $V_i=30$ km\, s$^{-1}$, $\Delta V=33$ km\, s$^{-1}$, $\rho_T=2.7$, $\Delta m_{\rho>\rho_T}=-0.86$. The red curve represents an exponentially damped sine function (labelled by ``Exp") with a damping time of 12 times the oscillation period (5~min). The ``Gauss" stands for the initial Gaussian and later exponential damping profile (beige). Dashed-dotted curves in the same colour mark the envelopes of the three oscillatory signals. Panel b: characteristic time scales as a function of $\rho_i/\rho_e$ (black) and $V_i/R$ (red). $t_B$ is when the loop becomes fully turbulent and $t_G$ is when the mass of the loop core equals zero. From panels c to f, each demonstrates the influence of one specific parameter (see the legend title) on the oscillatory pattern by varying its values while others remained unchanged in the equation. Panels g--h show the dependence of $\Delta m_{\rho>\rho_T}$ and $\Xi$ on $\rho_i/\rho_e$ and $\rho_T$.%For a given loop length of 200~Mm, P=300~s, $C_k=1333 km/s$.
	}
	\label{fig:model_oscillation}
\end{figure}

%Characteristics
This nonlinear oscillation is an interference of two cosine waves with kink and Alfv\'en frequencies, respectively. In addition, their amplitudes vary in time: the amplitude of the kink wave decreases from the initial value (see the first part on the right-hand side of Eq.~\ref{eq:nl_amplitude}) while that of the Alfvén wave increases from zero (see the second part on the right-hand side of Eq.~\ref{eq:nl_amplitude}) over time. This leads to two characteristics of nonlinear damping: (1) time-varying damping rate and (2) period/frequency drift from the kink frequency towards the Alfv\'en frequency of the mixing layer.
The oscillatory pattern described by this formula (see an example in the blue curve of Fig.~\ref{fig:model_oscillation}) differs from the traditionally used exponentially damped sine function (red curve in Fig.~\ref{fig:model_oscillation}a) and also from the initial Gaussian followed by an exponential function \citep{2013A&A...551A..40P} (hereafter short for Gaussian-exponential, see the beige curve in Fig.~\ref{fig:model_oscillation}a). Over around three periods, the amplitude of the nonlinear oscillation decreases by 80\% by visual inspection.  
This is consistent with the strong decay observed in decaying kink oscillations. Focusing on the damping pattern, see its envelope outlined by the blue dashed-dotted curve, the nonlinear oscillation damps slowly in the first one and a half cycles and then suddenly decays quickly in the last cycle before the tube becomes fully turbulent. 
For comparison, the exponentially damped sine function has a constant damping rate. Also, turbulence-induced nonlinear oscillation is different from the Gaussian-exponential profile (beige), which has the first Gaussian and later exponential damping rate.
Another characteristic of the nonlinear oscillation is the oscillation frequency drift in the later stage, see how the blue curve in Fig.~\ref{fig:model_oscillation}a deviates from the exponential sine wave solution after 9~min in time. This indicates the dominance of the mixing layer to the oscillation properties at later times. This period drift is a unique signature distinct from the linear damping profiles previously considered. \cite{2014SoPh..289.1999R} also reports the frequency drift in nonlinear kink oscillation, in the context of weak nonlinearity in standing kink oscillations of a stratified flux tube. %Tikhonchuk + 1995: frequency drift in Alfven wave.
%(At the first 1-2 cycles, it resembles exponential decaying kink oscillation, later it deviates)

Among the ten parameters in Eq.~\ref{eq:nl_amplitude}, which ones are crucial in damping? The decay is characterised by the damping time scale $t_B$($t_G$) and the decrease in amplitude over time. As shown in Fig.~\ref{fig:model_oscillation}b, $t_B$ and $t_G$ increase with $\rho_i$ and decrease exponentially with increasing $V_i/R$. This is because $V_i/R$ is proportional to $B$ and $G$, meaning it inversely scales with the time scale associated with zero momentum/mass of the loop core.
Considering the overall pattern, a higher $V_i/R$ (Fig.~\ref{fig:model_oscillation}c), a greater $C_1$ (Fig.~\ref{fig:model_oscillation}d), a lower density contrast (Fig.~\ref{fig:model_oscillation}e), and a smaller density threshold $\rho_T$ (Fig.~\ref{fig:model_oscillation}f) all contribute to faster damping. 
The first three factors result in a shorter damping time and a greater amplitude decrease simultaneously. In particular, $V_{i}$ is positively correlated with the magnitude of the amplitude.
As shown in panels e--f, the period drift becomes more pronounced with increasing $\rho_i$ or $\rho_T$ due to the rise in $\omega_A$ (greater discrepancy to $\omega_k$). %In addition, a greater $\omega_A$, (Fig.~\ref{fig:model_oscillation}f), leads to stronger damping in the later cycles.
As $\rho_T$ decreases, both the mass change rate $\Delta m_{\rho>\rho_T}$ and the mass per unit width $\Xi$ increases, see Fig.~\ref{fig:model_oscillation}g--h. As a result, the damping rate is higher in the early phase but slows down in the last cycle, see Fig.~\ref{fig:model_oscillation}f. 

After the entire tube becomes turbulent, the oscillation may continue decaying to a stationary stage. Such steady turbulence, when seen at low resolution, may resemble a decayless oscillation that has been discovered in recent years \citep{2012ApJ...751L..27W,2013A&A...552A..57N}, which is also evidenced in several numerical simulations \citep[e.g.][]{2016ApJ...830L..22A,2019ApJ...870...55G,2019FrASS...6...38K}. In this stage, the turbulence energy decays over time as $t^{-2}$, while the resulting internal energy increase remains minimal \citep{2024ApJ...966...68H}.

To recap, this nonlinear theory analytically describes the time variation of the mass and momentum of the loop resulting from the development of KHI, hereby determining the time series of the oscillation amplitude. The damping depends on multiple factors, including $V_i/R$ and density contrast. %We emphasize that this model can predict the oscillation amplitude for a set of given conditions, which is measurable in observations. 

\section[3]{Data Analysis Methods}\label{sec:methods}

\subsection[3.1]{Imaging data analysis}

We analysed two transverse loop oscillations on 30th May 2012 and 7th September 2017 selected from the 2019 Catalogue. These events are well observed by the Atmospheric Imaging Assembly (AIA, \citealt{2012SoPh..275...17L}) on board the Solar Dynamics Observatory (SDO, \citealt{2012SoPh..275....3P}) in the 171\,\AA\ channel. We downloaded the AIA 171\,\AA\ level 1 images from the JSOC centre. 

Time--distance analysis is applied to the image sequence to reveal the transverse oscillation as follows. First, a slit with a width of 5 pixels is put across the oscillating feature along the displacement direction and is used to make time--distance map. After that, the oscillatory signals are traced as the displacement of the loop boundary. The location of the loop boundary at each instant of time is determined by fitting the transverse intensity profile with a Gaussian. The full width at half maximum of the Gaussian is taken as the loop radius. The best Gaussian fit also gives the amplitude $I_i$ and background intensity $I_e$. Under the optically thin assumption, the ratio of their contrast can approximate the density contrast $\chi=\rho_e/\rho_i$ by fitting $I_e/I_i=l_e(\chi \rho_i)^2$ and $(l_e-1)(\chi\rho_i)^2+\rho_i^2=1$, where $l_e$ is the column depth of the background plasma along the line of sight, which is normalised by the column depth of loop, i.e., loop radius $R$. The search for the best-fitted $\chi$ for a given set of $I_0$ and $I_e$ is computed via Bayesian inference \citep[see details in Section~3.3 in][]{2019ApJ...884L..40A}. %by $\rho_e/\rho_i\approx\sqrt{1/(l_e(I_0/I_e)+1)}$, where $l_e$ is the column depth of the background plasma along the line of sight, which is normalised by the column depth of loop, i.e., loop radius $R$. Given that $l_e$ is usually much greater than the loop radius, we take $l_e=30$ for rough estimation in this work.
With the signals extracted from the time--distance map, the trend of the signals is obtained by smoothing with a window longer than the guessed period, and then the trend is subtracted to remove the background effect.
To measure the damping time $\tau$ in the first oscillation cycle, we fit the envelope of oscillatory signals with ${\rm exp}(-t/\tau)$. The maximum displacement amplitude $\xi_0$ of the signals is measured and then passed to calculate the velocity amplitude $V_{{i}}=2\pi\xi_0/{P}$ under the assumption of linear polarisation, where $P$ is the oscillation period obtained by fitting with the time series model, such as the exponential sine function.

\subsection{Bayesian parameter inference and model comparison}

In this study, we confront the imaging observations of transverse oscillations and our nonlinear model with the use of Bayesian analysis for parameter inference and model comparison. In the Bayesian framework, the posterior probability of a set of parameters $\theta = \{\theta_1, \theta_2, \ldots, \theta_i\, \ldots, \theta_N\}$ of a model $M$, conditional on observed data $D$, is given by Bayes' theorem $p(\theta|D)=p(D|\theta)p(\theta)/p(D)$. Here, $p(\theta)$ is the prior probability, which represents our belief about $\theta$ before considering the observations; $p(D|\theta)$ is the likelihood function, which gives the likelihood of obtaining a given data realisation $D$ that was observed as a function of the parameter vector; and $p(D)$ is the evidence. The posterior $p(\theta|D)$ is the updated belief in the possible values of $\theta$ after considering the data $D$.

Because of the high-dimensional parameter space of our model, with 6 independent parameters, we use Markov Chain Monte Carlo (MCMC) methods to approximate the posterior by random sampling in a probability space. In particular, we employed the Solar Bayesian Analysis Toolkit (SoBAT, \citealt{2021ApJS..252...11A}). The tool requires first to define the parameter space and priors over each parameter. Second, to provide a set of initial guesses of parameters to start the random walk via MCMC methods in the parameter space, and pass the samples to evaluate the goodness of fit to the data by the likelihood function. This step is repeated until the required number of good fits is reached. Then, application of statistical analysis to the obtained parameter samples is used to generate a histogram for each parameter, which are then used to approximate the posterior probability distribution function (PDF) of the corresponding parameter.  
The best estimate or best fit of the free parameters is identified as the point in parameter space where the global maximum posterior probability is reached. Traditional fitting approaches, such as the maximum likelihood estimate or the least-squares estimate, find the best fit at the global maximum without considering the uncertainties of parameters. The Bayesian approach enables us to compute in a consistent way the probability distribution of the parameters, conditional on the observed data, and the associated uncertainties.  The process of finding the optimal fit within the parameter space using MCMC sampling, is computationally efficient for relatively complex models with a substantial number of parameters, such as our nonlinear model. 
%\textcolor{magenta}{IA: perhaps we could keep this info. Concerning the computation time of MCMC, fitting 200 data points with $10^5$ samples takes around 30 min using a 12-core desktop, and fitting with 100 data points takes around 20 min. Taking $10^6$ samples will cost around one day.}

The method also allows quantitative model comparisons. The Bayesian evidence $p(D)$, also called the marginal likelihood, plays a relevant role in the Bayesian model comparison. This is the integral of the joint likelihood $p(D,\theta)=p(D|\theta)p(\theta)$ across the parameter space: $p(D)=\int_\theta p(D|\theta)p(\theta)d\theta$ and represents the probability of the observed data given a specific model. The integral over the full prior space penalises model complexity unless there is sufficient evidence in the likelihood function. This avoids the risk of overfitting of traditional approaches.
The absolute value of the evidence alone is not particularly useful; rather, the ratio of evidence between two competing models, known as the Bayes factor ($BF)$, quantifies the relative confidence in one model over the other. According to Kass and Raftery's scale \citep{1995Kass}, the Bayes factor -- defined as twice the logarithm of the ratio of two Bayesian evidence values -- can be interpreted as follows: $BF<2$: not worth mentioning, $BF\in 2-6$: positive preference, $BF\in 6-10$: strong preference, $BF>10$: very strong preference.

In our study, two methods are used to calculate Bayes factors. In Section~\ref{sec:fitting}, the Bayesian evidence for each considered model is computed from the samples derived from the MCMC fitting procedure. In Section~\ref{sec:predication}, the Bayes factor for each considered model is computed directly using the above definition of marginal likelihood, computing the integrals over a grid of points, and taking the corresponding ratios.

%For the MCMC fitting by Eq.~\ref{eq:nl_fitting}, the prior distributions of each parameter are set to be uniform with the range given based on empirical observations, e.g. $\rho_i\in[1,10]$. An example script is available online \footnote{Please see \url{https://github.com/Sihui-Zhong/nonlinear-kink-oscillations.git}}. %[The random walking iteration is set to be $10^5$ samples. 95\% credible intervals.]

%Concerning the computation time of MCMC fitting. Fitting 200 data points with $10^5$ samples takes around 30 min using a 12-core desktop, and fitting with 100 data points takes around 20 min. Taking $10^6$ samples will cost around one day. 

\section{Model validation}\label{sec:validation}

The validation of the model described in Section~\ref{sec:formula} is performed by fitting the nonlinear time series to data (see Section~\ref{sec:fitting}), and conducting a Bayesian model comparison between the proposed nonlinear model and the linear model with respect to the damping ratio (see Section~\ref{sec:predication}).

\subsection{Curve fitting}\label{sec:fitting}
We reconstruct the nonlinear turbulence damping function described in Section~\ref{sec:formula} for curve fitting as follows. %Considering describing a time series segment, $t_0$ is introduced to set the time starting the oscillation. 
To represent a time series segment, we introduce $t_0$ to mark the starting time of the oscillation. All density-related parameters are normalised by $\rho_e$, and we re-define $\Xi = \int_{y'(\rho_T/\rho_e)}^{y'(\rho_i/\rho_e)} \rho/\rho_e dy'$ to account for the effect of $\rho_T$ in computing $\overline{p}$.  
Since $C1$ is always tied with $R$ in the form of $C_1/R$, we take $C_1/R$ as one parameter.
A set of numerical simulations demonstrated in Appendix~\ref{sec:reduction} reveals $\Delta V$ scales linearly with $V_i$ by a factor of 1.1, thus we take $\Delta V = 1.1V_i$. To further reduce the number of free parameters, we consider the dependencies between certain parameters based on mixing theory. For example, the Alfv\'en frequency $\omega_{\rm A}$ in the mixing layer can be inferred from $\omega_{\rm k}$ and $\rho_i/\rho_e$. As demonstrated in \citet{2024ApJ...966...68H}, the mean density ${\rho_{\rm L}}$ in the mixing layer can be determined as the average of the density profile there, leading to $\omega_{\rm A} \propto 1/\sqrt{\rho_{\rm L}}$. However, direct computation of the density profile is computationally expensive. To improve the efficiency, we adopt a pre-computed look-up table of $\rho_{\rm L}$ as a function of $\rho_i$. Alternatively, $\rho_{\rm L}$ can be approximated as $\sqrt{\rho_i\rho_e}$ \citep{2019ApJ...885..101H}, providing a convenient and reasonably accurate estimate. With $\omega_{\rm k}\propto \sqrt{2/(\rho_i+\rho_e)}$, the ratio of Alfv\'en and the kink period can be estimated as 
\begin{equation}
 \frac{P_{\rm k}}{P_{\rm A}}=\frac{\omega_{\rm A}}{\omega_{\rm k}}\approx \frac{\sqrt{\rho_i/\rho_e+1}}{\sqrt{2 \rho_{\rm L}/\rho_e}}.    
\end{equation}
Here we use $P_{\rm k}/P_{\rm A}$ instead of $\omega_{\rm k}/\omega_{\rm A}$ as period is more straightforward in time-series analysis. %For $1<\rho_i/\rho_e<100$, $1<{P_{\rm k}}/{P_{\rm A}}<2.5$, meaning that $P_{\rm A}$ is always smaller than $P_{\rm k}$. 
Furthermore, the upper limit of $\Xi$ is $\sqrt{\rho_i/\rho_e}$, and the combination of $\rho_i, \rho_e$, and $\Xi$ relates to $\Delta m$ for a given density threshold $\rho_T$, as illustrated in Fig.~\ref{fig:model_oscillation}g--h.
To capture the role of $\rho_T$ in estimating $\Xi$, we introduce $\eta$, a percentage of the upper limit of $\Xi$, and $\Xi = \eta\sqrt{\rho_i/\rho_e}$. A smaller $\rho_T$ corresponds to a greater $\eta$. The interrelation between $\eta,\Xi,\Delta m$, and $\rho_i/\rho_e$ are given by
\begin{align}
%    \Xi &= \eta\textcolor{red}{\sqrt{\rho_i/\rho_e}},\\
    \Delta m &= \Xi - \rho_i\frac{1}{\sqrt{\rho_i/\rho_e}+1}.
\end{align}
Therefore, $P_{\rm A}$ and $\Delta m$ are dependent variables, reducing the number of parameters from 9 to \textbf{6}. The final formula is then expressed as
\begin{equation}\label{eq:nl_fitting}
    %\xi_{\rho>\rho_T} (t) = F(t,R,\rho_i/\rho_e,P_{\rm k}, P_{\rm A}, V_i, C_1, \Delta m, \Xi,\Delta V = 1.1V_{i},t_0),
    \xi_{\rho>\rho_T} (t) = F(t,t_0,{C_1/R},\rho_i/\rho_e,P_{\rm k}, V_i, \eta, P_{\rm A}=P_{\rm k}\frac{\sqrt{2\rho_{\rm L}/\rho_e}}{\sqrt{\rho_i/\rho_e+1}},\Delta V = 1.1V_{i},\Xi = \eta{\sqrt{\rho_i/\rho_e}},\Delta m=\Xi -\frac{\rho_i}{\sqrt{\rho_i/\rho_e}+1}),
\end{equation}
\noindent where $F$ is a function described in Eq.~\ref{eq:nl_amplitude}. 
%\textcolor{red}{Here we use $P_{\rm k}/P_{\rm A}$ instead of $\omega_{\rm k}/\omega_{\rm A}$ as period is more straightforward in time-series analysis. }

To fit the time series by our nonlinear model with 6 parameters, we use the MCMC fitting method within Bayesian inference \citep{2021ApJS..252...11A}. For the MCMC fitting by Eq.~\ref{eq:nl_fitting}, the prior distributions of parameters are set to be uniform with the range given based on empirical observations, e.g. $\rho_i/\rho_e\in[1,10]$. %, while $C_1$ is assigned to a normal distribution with the mean of 0.3 \citep{2024ApJ...966...68H}. 
An example script is available online \footnote{Please see \url{10.5281/zenodo.15639588}. Note that displacement amplitudes are given in units of Mm instead of km to avoid arithmetic errors in the calculation of Bayesian evidence}. %\url{https://github.com/Sihui-Zhong/nonlinear-kink-oscillations.git}

To perform a model comparison between the nonlinear and linear damping models, the same data is fitted with the exponential damped sine function 
\begin{equation}\label{eq:exp}
    \xi=\xi_0\exp{(-\frac{t-t_0}{\tau_e})}\sin(\frac{2\pi (t-t_0)}{P_e}+\phi),
\end{equation} 
\noindent where $\tau_e$ is the damping time. For the oscillation period, we use a subscript to indicate the specific model: $P_e$ corresponds to the period in the exponential damping model, $P_g$ represents the period in the Gaussian-exponential model (Eq.~\ref{eq:gauss}). The Bayesian evidence of each function and their ratio, i.e., the Bayes factor, is calculated to quantify how much better one of the two competing models explains the data. The exponential damping function is used because it is linear and commonly applied to approximate the decay pattern in decaying kink oscillation events.
Another linear damping function applied to kink oscillations is the initially Gaussian decay followed by an exponential function \citep{2013A&A...551A..40P}:
\begin{equation}\label{eq:gauss}
    \xi = \begin{cases}
        \xi_0 \exp{-\frac{(t-t_0)^2}{2\tau_g^2}}\sin(\frac{2\pi (t-t_0)}{P_g}+\phi) & \text{$ t \leq t_s$,}\\
        \xi(t_s)\exp{(-\frac{t-t_s}{\tau_e})}\sin(\frac{2\pi (t-t_0)}{P_g}+\phi) & \text{$t > t_s, t_s=\tau_g^2/\tau_e$}.
    \end{cases}     
\end{equation}
\noindent Here $\tau_g, \tau_e$ are the damping times for the Gaussian and exponential functions, respectively. This function will be used in Section~\ref{sec:obs}.
The comparison between a linear model and our nonlinear model can serve to: (1) check the ability of the method to distinguish between nonlinear and linear oscillation; and (2) re-evaluate the accuracy of the measurement of oscillation properties, especially the period, which is an important input in seismological applications.

\subsubsection{Fitting with observational data}\label{sec:obs}
As a preliminary step, curve fitting by the model time series is first tested on synthetic data before being applied to real observations. As demonstrated in Appendix~\ref{appendix:curvefit}, the fitting procedure can recover the input parameters and distinguish the nonlinear and linear models even when the signals are contaminated with noise. Also the posterior distribution of key parameters, such as $\rho_i,V_i,P_{\rm k}$, are well-constrained. This indicates that the MCMC fitting with 6 parameters is reliable.
The same fitting procedure and model comparison are applied to real data.
A typical event is selected from the 2019 Catalogue. On 7th September 2017, an M-class flare excited a decaying kink oscillation of a bundle of nearby coronal loops. The SDO/AIA well captured this event. We analysed this event using AIA 171\,\AA\ level 1 images implemented with typical time--distance analysis (see details in Section~\ref{sec:methods}). 

\begin{figure}[ht]
	\centering
	\includegraphics[width=\linewidth]{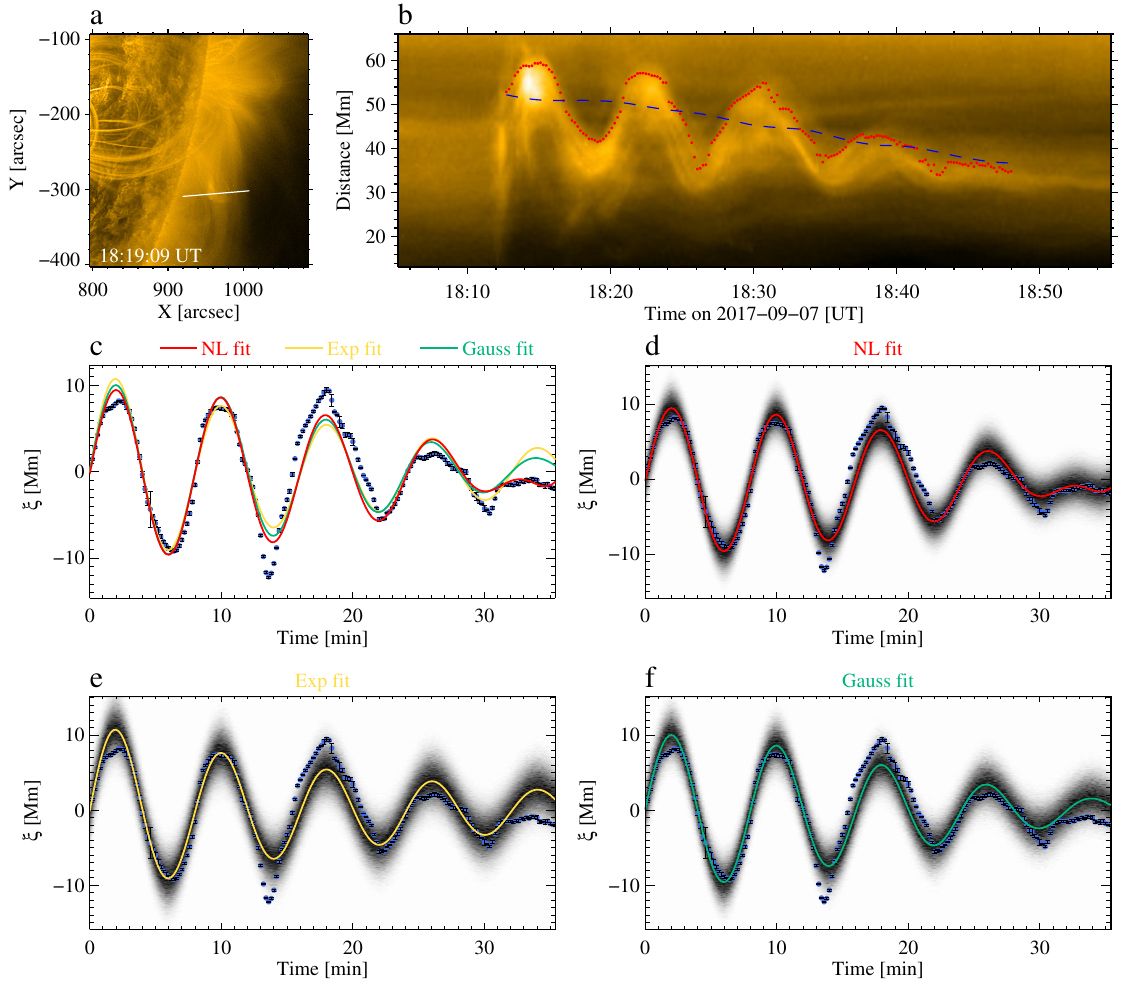}
	\caption{
        Fitting the observed decaying kink oscillation from 18:10~UT to 18:50~UT on 7th September 2017 with three models: the nonlinear turbulence damping model (Eq.~\ref{eq:nl_fitting}, labelled NL), exponential damping model (Eq.~\ref{eq:exp}, labelled Exp), Gaussian-exponential damping model (Eq.~\ref{eq:gauss}, labelled Gauss). Panel a: SDO/AIA 171\,\AA\ image showing the oscillating plasma structure off the solar limb. The white slit was put across the targeted structure and along the oscillating direction and used to make time--distance map in panel b. Panel b: The oscillatory pattern of the loop studied, traced by the red dots. The blue dashed curve is the trend of the signals. Panel c--d: Best fit of nonlinear function (red, d), exponential damping function (yellow, e), and Gaussian-exponential damping profile (green, f) with the observed oscillation (blue dots with error bars). The best-fitted period and damping time of each function can be found in Table~\ref{table:BF_obs}. The background shading is the posterior predictive distribution, i.e., the distribution of possible unobserved values conditional on the observed values, with darker colour indicating a higher possibility.
	}
	\label{fig:fit_obs}
\end{figure}

Fig.~\ref{fig:fit_obs}a shows the image of the oscillating structure (marked by the white slit) that is off the solar limb. The transverse oscillatory pattern is revealed in the time--distance map as shown in Fig.~\ref{fig:fit_obs}b, where the oscillatory signals (indicated by the red dots) and its trend (blue dashed curve) are extracted.
Then the detrended signals are fitted by Eq.~\ref{eq:nl_fitting}, Eq.~\ref{eq:exp} and Eq.~\ref{eq:gauss} respectively, based on which the Bayesian model comparison is done. In addition, from Fig.~\ref{fig:fit_obs}b, the visually estimated loop diameter is around 10~Mm and the oscillation period is within $7-9$~min.

\begin{figure}[ht]
	\centering
	\includegraphics[width=\linewidth]{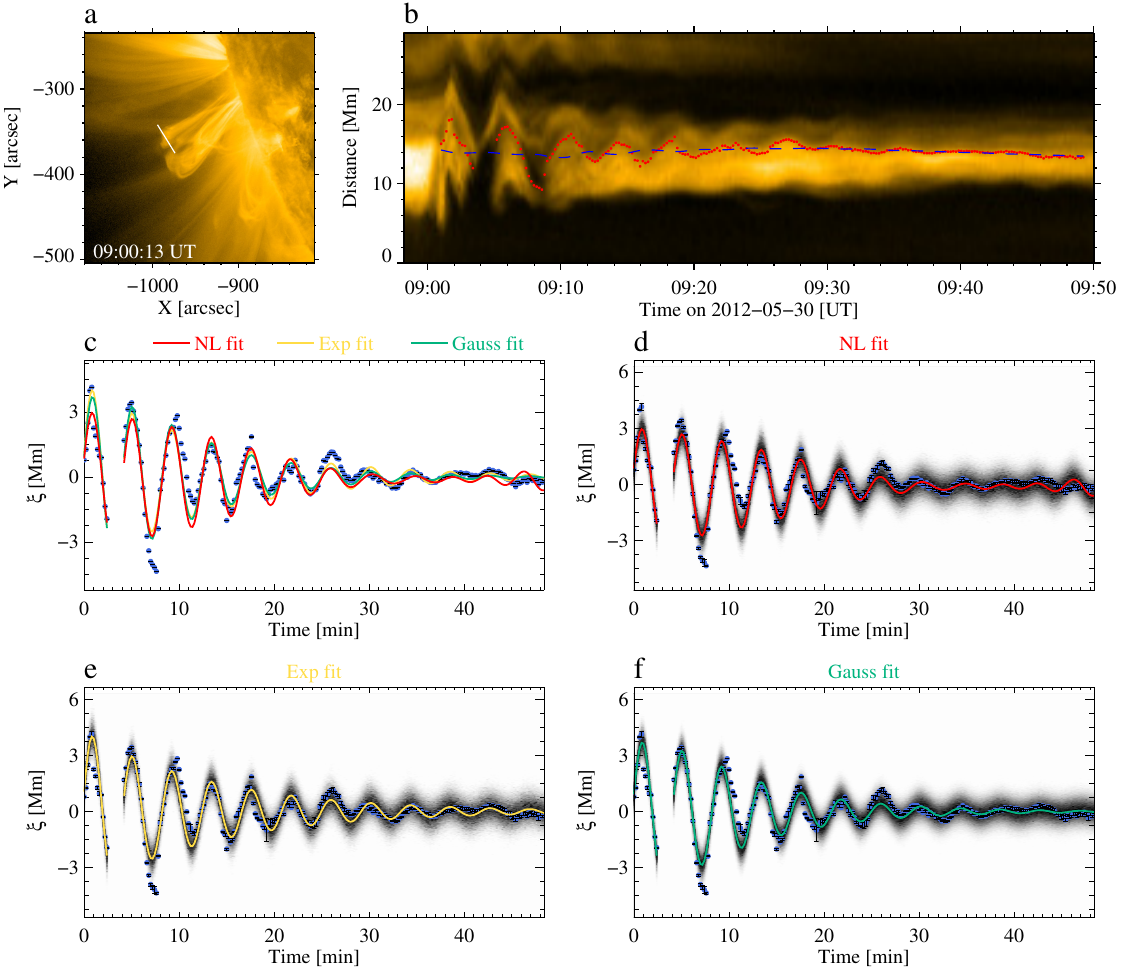}
	\caption{
        Similar to Fig.~\ref{fig:fit_obs} but for decaying kink oscillation during 09:00~UT to 09:50~UT on 30th May 2012 with three models. Signals between around 09:03~UT and 09:05~UT are missing.
	}
	\label{fig:fit_obs_2012}
\end{figure}

The best-fitted parameters of our nonlinear function are displayed in the penultimate row of Table~\ref{table:pars}. If taking $C_1=0.3$, the loop radius of $7.5^{+2.5}_{-1.5}$~Mm roughly agrees with the visually observed 5~Mm. 
The period of $482^{+5}_{-4}$~s is consistent with previous measurement (8.32~min) in the 2019 Catalogue. With $\rho_i/\rho_e=5.1$ and $\eta=0.21$, it gives the density threshold to define the lower limit of the loop core $\rho_T = 81\% \rho_i$. As seen in Fig.~\ref{fig:fit_obs}c, the best fit of the nonlinear (NL) function (red) and posterior predictive distribution for the amplitude (grayscale shading) match the observed signals (blue dots). The best exponential (Exp) fit is indicated by the yellow curve in Fig.~\ref{fig:fit_obs}d, and the best Gaussian-exponential (Gauss) fit is depicted by the green curve in Fig.~\ref{fig:fit_obs}f. The best-fitted key parameters of each fit are shown in Table~\ref{table:BF_obs}. %The best-fitted parameter for the Gauss fit is: $\xi_0=10$~Mm,$\tau_g = 1003$~s,$\tau_d=500$~s,$P=482$~s,$t_0=98$,$\phi=1.3$. 
The nonlinear fit is better than the Exp and Gauss fits, particularly at the first crest and the last oscillation cycle.
As displayed in Table~\ref{table:BF_obs}, the Bayes factor for the nonlinear model over the Exp model $BF_{\rm NL/Exp}$ is 86.7, yielding very strong evidence in favour of the nonlinear wave damping model. Comparing the nonlinear and Gauss profiles, the Bayes factor of 43.6 gives decisive evidence in favour of the nonlinear model.
This is consistent with the analysis result on the same event in \citet{2023MNRAS.525.5033Z}. 
The same analysis was additionally applied to the decaying kink oscillation event on 30th May 2012 
(see Fig.~\ref{fig:fit_obs_2012}). Given that $BF_{\rm Exp/NL}=$ 25.3 and $BF_{\rm Gauss/Exp}=$23.8 (see Table~\ref{table:BF_obs}), very strong evidence for the Gaussian-exponential damping model is obtained for this event. 
%\textbf{Including the first five-minutes data leading to $BF_{\rm EXP/NL}=32$ and $BF_{\rm Gauss/Exp}=74$, yielding very strong evidence for the Gaussian-exponential model, and the linear model remains preferred in this case.}
When comparing the inferred parameters for both events, we note that the first analysed event, for which the evidence in favour of the nonlinear model is very strong, has higher values of $V_i$ and displacement amplitude (2 times higher), these parameters are directly related to nonlinearity. Overall, these results show that the Bayesian model comparison method can be applied to more observed events to assess and quantify the nonlinearity of their oscillatory dynamics.

\begin{table}[ht]
\centering
\begin{tabular}{lcccccc} 
 \hline
  Event & $C_1/R$ [Mm$^{-1}$] & $\rho_i/\rho_e$ & $V_i$ [km\,s$^{-1}$] & $P_{\rm k}$ [s] &  $\eta$ & $t_0$ [s] \\
  \hline
  2017-09-07 & $0.04^{+0.01}_{-0.01}$ & $5.2^{+0.5}_{-0.6}$ & $128^{+8}_{-9}$ & $482^{+5}_{-4}$  & $0.21^{+0.07}_{-0.05}$ & $-1^{+7}_{-7}$  \\
\hline
  %2012-05-30 & $0.04^{+0.01}_{-0.01}$ & $3.1^{+0.2}_{-0.1}$ & $93^{+8}_{-7}$ & $254^{+3}_{-3}$ & $0.99^{+0.01}_{-0.1}$  & $-7^{+5}_{-5}$  \\
  2012-05-30 & $0.07^{+0.01}_{-0.01}$ & $3.0^{+0.1}_{-0.1}$ & $77^{+5}_{-6}$ & $255^{+2}_{-2}$ & $1.00^{+0.00}_{-0.13}$  & $-13^{+5}_{-5}$  \\
  \hline
\end{tabular}
\caption{The best-fitted parameters obtained by MCMC fitting with Eq.~\ref{eq:nl_fitting} to two selected events. The errors are determined by the 95\% credible intervals.}
\label{table:pars}
\end{table}

\begin{table}[ht]
\centering
\begin{tabular}{ccccccccccc} 
 \hline
  & $\mathbf{M}_{\rm NL}$ & \multicolumn{2}{c}{$\mathbf{M}_{\rm Exp}$} & \multicolumn{3}{c}{$\mathbf{M}_{\rm Gauss}$} &  &  &  \\\cline{2-7}
 Event & $P_{\rm k}$ [s] & $P_e$ [s] & $\tau_e$ [s] & $P_g$ [s] & $\tau_g$ [s] & $\tau_e$ [s] & $BF_{\rm NL/Exp}$ & $BF_{\rm NL/Gauss}$ & $BF_{\rm Gauss/Exp}$ & For \\
 \hline
 2017-09-07 & $482^{+5}_{-4}$ & $482^{+9}_{-5}$ & $1418^{+313}_{-200}$ & $481^{+6}_{-4}$ & $1055^{+112}_{-68}$ & $454^{+299}_{-398}$ & 86.7 &  43.1 & 43.6 & $\mathbf{M}_{\rm NL}$\\
% 2012-05-30 & $254^{+3}_{-3}$ & $250^{+2}_{-2}$ & $695^{+92}_{-81}$ & $250^{+2}_{-2}$ & $122^{+35}_{-20}$ & $443^{+68}_{-53}$ & -32  & -106 \ & 74 & $\mathbf{M}_{\rm Gauss}$\\
  2012-05-30 & $251^{+2}_{-2}$ & $251^{+2}_{-2}$ & $815^{+86}_{-96}$ & $251^{+2}_{-2}$ & $605^{+173}_{-29}$ & $576^{+130}_{-150}$ & -25.3 &  -49.1 & 23.80 & $\mathbf{M}_{\rm Gauss}$\\
 \hline
\end{tabular}
\caption{Bayesian model comparison between nonlinear turbulence damping model ($\mathbf{M}_{\rm NL}$), exponential model ($\mathbf{M}_{\rm Exp}$) and Gaussian model ($\mathbf{M}_{\rm Gauss}$) for observed events. $P/\tau$ is the best-fitted oscillation period/damping time for each model, with errors determined at their 95\% credible intervals. $BF$ is the Bayes factor of two competing models denoted by the subscript.}
\label{table:BF_obs}
\end{table}

\subsection{Bayesian model comparison with damping rate}\label{sec:predication}

%The driving mechanism acting behind the decaying kink oscillations of coronal loops can be revealed by the comparison between theoretically predicted and empirical dependence of observables, e.g., the linear scaling of $\tau$ and $P$ justifies the resonant absorption mechanism. % 
In the previous analysis, we assessed the relative plausibility of each damping model in explaining the time evolution of the oscillation amplitude for both synthetic data and real data. We now apply Bayesian model comparison to the damping ratio, a measure describing how rapidly the oscillations decay. This can give additional information on the physical mechanism behind the decaying kink oscillations of coronal loops.
In the Bayesian framework, the Bayesian evidence or marginal likelihood represents the joint probability of data and model over the whole parameter space. More importantly, the ratio of marginal likelihoods allows pairwise model comparison.
Here, we quantify the evidence of the nonlinear turbulence damping model and the linear resonant absorption (RA) model for the decaying kink oscillation of coronal loops, on the basis of their predictions of the damping rate. %the probability/goodness of the model fitting the observational data, using the Bayesian evidence. 

Before computing the Bayesian evidence for the nonlinear model, it is necessary to reduce its complexity by setting less important parameters as constants and by taking advantage of the known correlations arising from the mixing theory, simulations, and observations. %Note that only $V_{i}/R$ can change the maximum displacement among the ten parameters.
In previous hydrodynamic simulations, $C_1$ is within the range of 0.3-0.5 \citep{1974JFM....64..775B,2023MNRAS.520.1738H}, and in 3D-like simulations, it is 0.3 \citep{2020JFM...900A..16B,2024ApJ...966...68H}, so we set its default value as $C_1=0.3$.  
In EUV imaging observations, the 171\,\AA\ channel is commonly used in observing transverse oscillations in coronal loops, as the loops are seen in high contrast in 171\,\AA\ wavelength. In addition, as reported by \citet{2016ApJ...830L..22A,2017ApJ...836..219A}, Fe IX 171\,\AA\ is a core line that captures the loop core's dynamics. Therefore, plasma observed in 171\,\AA\ images can be considered to be confined by a high-density threshold close to $\rho_i/\rho_e$, say $95\%\rho_i/\rho_e$.
As mentioned in Sections~\ref{sec:formula} and \ref{sec:fitting}, the mixing theory \citep{2019ApJ...885..101H} can predict the mass change rate ($\Delta m_{\rho>\rho_T}$), the mass per unit width $\Xi$, and $\omega_{\rm A}$ for a given set of $\rho_i/\rho_e, \rho_T$, and $\omega_{\rm k}$. 
As a result, we can reduce the number of parameters to four: $R, \rho_i/\rho_e, \omega_{\rm k}$ (or $P_{\rm k}), V_i$, with 4 default constants: $C_1=0.3, \rho_e=1, \Delta V=1.1V_i, \rho_T=0.95\rho_i/\rho_e$. In the context of our 6-parameter nonlinear function (Eq.~\ref{eq:nl_fitting}), $\rho_T=0.95\rho_i/\rho_e$ corresponds to $\eta<0.1$, with its exact value depends on $\rho_i/\rho_e$. And $t_0$ is set as zero since it does not affect the damping rate.

\begin{figure}[ht]
	\centering
    \includegraphics[width=\linewidth]{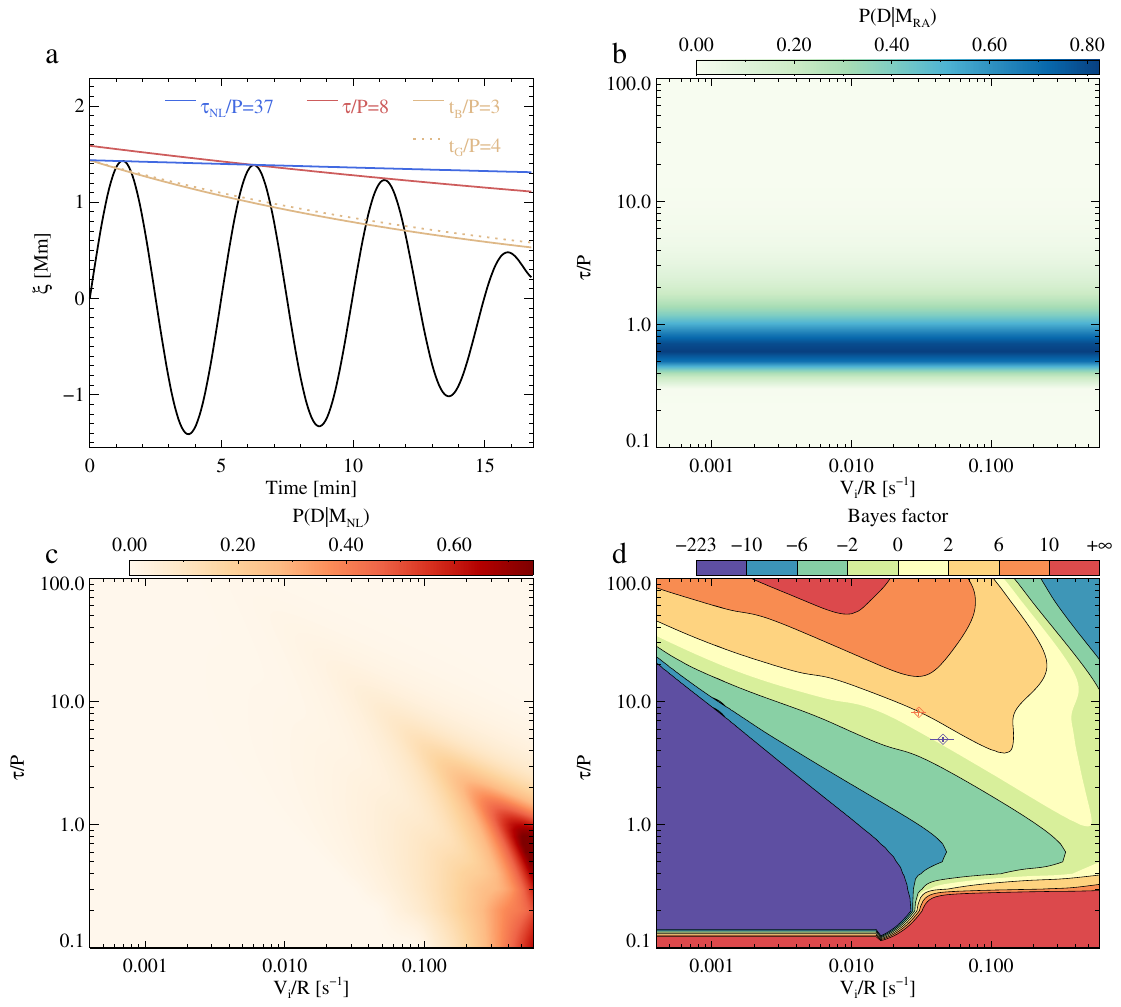}
	\caption{
        Bayesian model comparison between nonlinear (NL) and RA models on the plane of quality factor ($\tau/P$) vs. $V_i/R$. Panel a: the characteristic damping time in the first (blue), second (red) cycle and overall (beige) cycles. Panel b--c: marginal likelihood for the RA and nonlinear models. Panel d: Bayes factor distribution, B$_{\rm NL/RA}$. The black contours indicate Bayes factor levels of -10, -6, -2, 2, 6, and 10, supporting the RA (the first three) or the nonlinear model (the latter three). The orange and purple diamonds indicate the position in the plane for the two analysed oscillation events on 7th September 2017 and 30th May 2012. The input data for these two events are measured in imaging data (see Section~\ref{sec:methods}): for the 2012-05-30 event, $\tau/P=4.9\pm0.3, R=2.3$~Mm, $V_i={106\pm20}$~km\,s$^{-1}$, and for the 2017-09-07 event, $\tau/P=8.2\pm0.5, R=4.6$~Mm, $V_i={141\pm17}$~km\,s$^{-1}$. 
	}
	\label{fig:bf}
\end{figure}

In the condition of a high $\rho_T$ (or low $\eta$), the instantaneous damping rate of nonlinear oscillation is low at first and becomes much higher in later times, see Fig.~\ref{fig:model_oscillation}a. In other words, the characteristic damping time varies over time. In the first oscillation cycle, the role of the mixing turbulent layer in controlling the oscillation is negligible when $\rho_T\rightarrow\rho_i$, which is virtually equivalent to setting $D=0$ in Eq.~\ref{eq:velocity_amp}-\ref{eq:nl_amplitude}. In this case, the loop oscillation is well represented by the oscillation of its core.
Consider the oscillation in the loop core only, the oscillation profile is determined by the ratio $\frac{1-Bt}{1+Gt}$. Thus, we can approximate the damping profile of the core momentum and core mass by an exponential function with damping times of $t_B = 1/B$ and $t_G=-1/G$, respectively.
Therefore, the nonlinear damping time of the loop oscillation in the first cycle is estimated by $\tau_{\rm NL} = \frac{t_Bt_G}{t_G-t_B}$, %\textcolor{red}{Here $\tau_{NL} $ should not be confused with the nonlinear time, which is the timescale over which fluctuations at different scales interact and transfer energy in the turbulence cascade.} 
and the quality factor for the nonlinear model is given by %[\textcolor{magenta}{note that I put rm on all \textbf{nonlinear}subscripts]}
\begin{equation}\label{eq:tau_nl}
    \frac{\tau_{\rm NL}}{P} = \frac{t_Bt_G}{t_G-t_B}\frac{1}{P}= \frac{\pi}{2}\frac{R}{\Delta V C_1}\frac{\sqrt{\rho_i}+\sqrt{\rho_e}}{(\rho_i\rho_e)^{1/4}}\frac{\rho_i(\sqrt{\rho_i}+\sqrt{\rho_e})}{\rho_i\sqrt{\rho_e}+\Delta m_{\rho>\rho_T}(\sqrt{\rho_i}+\sqrt{\rho_e})}\frac{1}{P}.
\end{equation}
Take the oscillation displayed as the black curve in Fig.~\ref{fig:bf}a as an example, its period is $P=300$~s, the estimated $t_B = 1009.7$~s, $t_G=1111.5$~s, which gives $\tau_{\rm NL}=37P$. The exponential damped sine function with $\tau_{\rm NL}$ (blue) fits the envelope of the first oscillation cycle well, justifying the approximation of characteristic damping time in the first cycle.
The nonlinear-model quality factor of 37 is much higher than that given by $t_B$ ($t_G$) (see the beige curve), 3 (4). In the second cycle, see the red profile, the quality factor of 8 indicates a faster decay. Hence, the characteristic damping time in the first 1-2 cycles is significantly different from the typical damping time determined by the resonant absorption model, which is 2--6 times the period \citep{2019ApJS..241...31N}. This difference inspires us to consider Bayesian model comparison with respect to the quality factor, $\tau/P$, for the first oscillation cycle.

According to the linear resonant absorption model \citep{2002A&A...394L..39G}, the quality factor for the whole time series is given by 
\begin{equation}
    \frac{\tau}{P} = \frac{2}{\pi}\frac{R}{l}\frac{(\rho_i/\rho_e+1)}{(\rho_i/\rho_e-1)},
\end{equation}
\noindent where $l$ is the width of the transition layer of the oscillating loop. As this formula does not involve amplitude, resonant damping is independent of amplitude. On the contrary, the nonlinear damping rate is amplitude dependent, since $\Delta V/R$ is proportional to the initial wave amplitude. In the following, the Bayesian model comparison will be performed on the plane of observables $\tau/P$ and $V_i/R$.

For a given model $M_i$ with parameter vector $\theta$ proposed to explain the observed data $D$, the marginal likelihood is calculated by %\textcolor{magenta}{[I added a blank space between likelihood and prior and between prior and dtheta]}
\begin{equation}
    p(D|M_{i})=\int_\mathbf{\theta} p(D|\mathbf{\theta}, M_{i})\, p(\mathbf{\theta}|M_{i})\,d\mathbf{\theta}.
\end{equation}
\noindent For the nonlinear model, its parameter vector $\theta_{\rm NL}=\{V_i/R, P, \rho_i/\rho_e\}$, and for the RA model, $\theta_{\rm RA}=\{V_i/R, l/R, \rho_i/\rho_e\}$. To compare two models explaining the same data, the ratio of marginal likelihoods is used to estimate the Bayes factor: 
\begin{equation}
    BF_{\rm NL/RA}=2\ln{\frac{p(D|M_{\rm NL})}{p(D|M_{\rm RA})}}=2\ln{\frac{\int_\mathbf{\theta} p(D|\mathbf{\theta_{\rm NL}}, M_{\rm NL})\, p(\mathbf{\theta_{\rm NL}}|M_{\rm NL})\,d\mathbf{\theta_{\rm NL}}}{\int_\mathbf{\theta} p(D|\mathbf{\theta_{\rm RA}}, M_{\rm RA})\, p(\mathbf{\theta_{\rm RA}}|M_{\rm RA})\,d\mathbf{\theta}_{\rm RA}}}.
\end{equation}
%[\textcolor{magenta}{note that the parameter vector for each of these models is different, so we need to introduce $\theta_{\rm NL}$ and $\theta_{\rm RA}$ in this equation already]}

We numerically calculate the marginal likelihood over a grid of points in data space $D=\{V_i/R,\tau/P\}$ with $203\times 241$ points, with parameter vector $\theta_{\rm NL}$ for $M_{\rm NL}$ and $\theta_{\rm RA}$ for $M_{\rm RA}$. %we make synthetic data space for each model: $D=(\tau/P,\theta_{\rm NL}=\{V_i/R, P, \rho_i\})$ for the \textbf{nonlinear}model and $D = (\tau/P, \theta_{\rm RA}=\{V_i/R, l/R, \rho_i/\rho_e\})$ for the RA model.
The range of parameters is set according to all published kink oscillations of coronal loops, including the 2019 Catalogue and related literature  \citep[e.g.,][]{2023NatSR..1312963Z}: $V_i/R \in [0.0004, 0.6]$~Hz ($V_i \in [1, 310]$~km\,s$^{-1}$, $R \in [0.5, 12.5]$~Mm), $P \in [90, 1490]$~s, $\rho_i/\rho_e \in [1.1,10]$ with $\rho_e=1$ as default. The grid size for each parameter is constrained by limitations in observations, e.g., the interval for length is 0.5~Mm, which is about the AIA image resolution; for $P$ it is 12~s, around the AIA temporal cadence. 
As the time scale in the nonlinear model is determined by $V_i/R, P$, and $\rho_i/\rho_e$, we combine $V_i$ and $R$ as a single parameter. The range for the data is set as $\tau/P\in[0.1,100]$. For the RA model, we adopt $l/R \in [0.1,2], \rho_i/\rho_e\in[1.1,10]$ according to \cite{2021ApJ...915L..25A}. %$\xi\in [0.1, 30]$~Mm
We assume a Gaussian likelihood function $p(D|\mathbf{\theta}, M_{i}])$ for both models, given by
\begin{equation}
    p(D|\mathbf{\theta}, M_{i}) = \frac{1}{\sqrt{2\pi}\sigma} \exp{(-\frac{[\frac{\tau}{P}-\frac{\tau}{P}(\mathbf{\theta})|M_{i}]^2}{2\sigma^2})},
\end{equation}
\noindent where $\sigma$ is the uncertainty of quality factor, being set as $\sigma=0.2\frac{\tau}{P}$ according to the fitting results as shown in Table~\ref{table:BF_obs}. The priors are assumed to be uniform over the assumed parameter ranges.

%For the RA model, we compute the marginal likelihood by integrating the weighted likelihood over 2 dimensions ($\rho_i/\rho_e$, $R/l$) of parameter space on the surface of $\frac{\tau}{P}$ vs. $V_i/R$.
The resulting marginal likelihood on the surface of $\tau/P$ vs. $V_i/R$ for the RA model is shown in Fig.~\ref{fig:bf}b, where the magnitude indicates the rational evidence (plausibility) for a particular combination of $\tau/P$ and $V_i/R$ described by the RA model. The calculation marginalises over the model parameters, incorporating predictions from all possible considered combinations of $l/R$ and $\rho_i/\rho_e$.
As shown in Fig.~\ref{fig:bf}b, the marginal likelihood for the RA model is independent of amplitude as expected, since the model is linear and does not depend on $V_i/R$. Regions of high marginal likelihood for the RA model are enclosed within the range $\tau/P=0.3-2$. %\textcolor{magenta}{[There is something I don't understand well. In the distribution of marginal likelihood for RA, it seems that the marginal likelihood for values of td/p in the range between 1 and 10  increases for increasing value of td/P. But the marginal likelihood should decrease for increasing td/P, e.g., see Figure 1 (right) in Arregui (2021)].} 
The marginal likelihood for the nonlinear model is shown in Fig.~\ref{fig:bf}c. Regions of higher marginal likelihood form an elongated structure that extends from large amplitude values with relatively strong damping to lower amplitude regions with relatively weaker damping.
The elongated structure reflects the inverse relationship between quality factor and oscillation amplitude in the model. This structure of the evidence is also apparent in the diagonal distribution of the Bayes factor in Fig.~\ref{fig:bf}d.

Taking the ratio of the two marginal likelihood distributions in Figs.~\ref{fig:bf}b--c results in the complex Bayes factor distribution shown in Fig.~\ref{fig:bf}d. Different colours indicate varying levels of relative evidence supporting the nonlinear model over the RA model (or vice versa), according to the evidence classification scale by \citet{1995Kass}. This leads to distinct regions with different degrees of relative support for each model. Despite the complexity, valuable insights can be extracted. 
Regions with strong damping with $\tau/P<0.2$ prefer the nonlinear model with very strong evidence. As the damping weakens, the supports for the resonant damping extend into a triangular purple-blue region. A similar triangular purple-blue region, supporting the resonant damping model, appears in the upper right corner of the data space considered. From an observational perspective, of particular interest is the region with large oscillation amplitudes and strong damping combinations in which the nonlinear model is also preferred. The remaining diagonal band with strong or very strong evidence favouring the nonlinear model occurs in the weak damping regime at intermediate amplitude.
%For instance, in the low-oscillation amplitude regime ($V_i/R < 0.02$) the Bayes factor initially favours resonant damping in plane-parallel horizontal regions within the very strong damping regime \textcolor{red}{at $0.2<\tau/P<2$}. The remaining diagonal band contains two regions with strong or very strong evidence favouring the \textbf{nonlinear}model. One occurs in the weak damping regime at intermediate amplitude, while the other --more relevant from an observational perspective-- corresponds to large oscillation amplitudes and strong damping combinations. 

The 2D Bayes factor distribution in the ($\tau/P$, $V_i/R$)-plane (Fig.~\ref{fig:bf}d) provides a bird's-eye overview of the relative plausibility of the nonlinear and RA models for decaying kink oscillations over a given data-space, offering a valuable tool for understanding the general structure of the evidence. For given observed values of $V_i/R$ and damping ratio in the first cycle, this distribution serves as a broad reference for assessing the expected dominance of either model.
However, conclusions about the plausibility of each model from the position of individual events in the 2D space must be made carefully.  
For example, the positions of the 2012-05-30 event (marked by the purple circle) and the 2017-09-07 oscillation event (marked by the orange circle) lie in the region with equal support for both models $BF\in[-2,2]$, which disagrees with the model comparison result obtained from the time-series analysis. 
%[the position of the 2012-05-30 event (marked by the purple circle) lies in the region with positive support for the RA model, which agrees with the model comparison result obtained from the time-series analysis. In contrast, the 2017-09-07 oscillation event (marked by the orange circle), identified as plausibly nonlinear in Section~\ref{sec:obs} is located near the boundary of the same RA-favoured region.] 
The reason for this seeming discrepancy is that the evidence distribution in Fig.~\ref{fig:bf}d was computed by integration over large ranges of equally plausible values for the parameters, such as density contrast and period, while the previous event-specific evidence assessment was based on constrained parameters from the time-series analysis. 
Let us compute the Bayes factor for those two particular events by fixing measurable parameters such as $R,\rho_i/\rho_e,$ and $P$, while the remaining parameters retain their uniform priors within the same ranges in the 2D evidence distribution analysis. Following this approach, the 2017-09-07 event ($\tau/P=8.2, \rho_i/\rho_e=10.02, R=4.6$~Mm, $V_i=141$~km/s, $P=482$~s) yields a Bayes factor of 6.1, providing strong evidence in favour of the nonlinear model. For the 2012-05-30 event ($\tau/P=4.9, \rho_i/\rho_e=23.26, R=2.3$~Mm, $V_i=106$~km/s, $P=251$~s), the Bayes factor is negative infinity, strongly supporting the RA model.
Now the results from applying the model comparison to the time-series in Section~\ref{sec:obs} and the damping ratio in Section~\ref{sec:predication} are consistent. 
Note also that the measurements of density contrast carry significant uncertainty, as they are derived from the square root of internal-to-external intensity ratios. This reinforces the need for model comparisons using time-series signals, as in Section~\ref{sec:obs}, for a more definitive distinction between linearity and nonlinearity. Nevertheless, for broad predictive insights, the 2D Bayes factor map remains a valuable tool for assessing model plausibility based on the damping ratio.

%\textcolor{blue}{[comments: need emphasizing the value of the 2D Bayes factor distribution for general trends while clarifying that specific events require a more detailed posterior analysis]}

\section{Conclusion and Discussion}\label{sec:dis}

This work provides an analytic formula for the nonlinear transverse oscillation damped by KHI-induced turbulence. The damping is governed by 9 (or 6) parameters including $R, \rho_i/\rho_e$, and $V_i$. Some of them are correlated.
The characteristic damping time decreases as $V_i/R$ increases and the density contrast decreases. The amplitude-dependent damping is consistent with the parametric survey on velocity amplitude effects in oscillation damping by 3D MHD simulations \citep{2016A&A...595A..81M}.
The key observational signatures of the nonlinear damping oscillation are (1) uneven damping rate and (2) period drift.
The former visually manifests as a non-exponential decay profile. The period drift can be identified in real observation using wavelet analysis, provided that the number of oscillation cycles is sufficiently large \citep{2004SoPh..222..203D}, typically around 5 cycles. 
In our model, the damping rate is time-dependent because the turbulence takes time to develop. In general, nonlinear decaying oscillations initially decay slowly due to the minor role of turbulence with limited size. As turbulence grows becomes dominant, strong decay occurs, especially before the whole tube reaches full turbulence. This differs from classic amplitude-dependent nonlinear damping, as the damping here is governed by the time-evolving turbulence, not just instantaneous amplitude.

Our 6-parameter nonlinear function can be used to fit oscillatory signals via MCMC methods. The overall oscillation pattern inherently constrains the posterior distributions of $V_i, \rho_i/\rho_e, P_{\rm k}$, ensuring well-defined parameter estimates. The reliability of the best-fit model is confirmed through tests with both synthetic data and observed data. 
The validity of this theory is demonstrated by (1) a good match with the observed oscillations; (2) consistency between observed loop properties (e.g., loop width) and fitting result. %\sout{, and (3) agreement between internal correlations predicted by mixing theory and those revealed in the MCMC sampling.}

%The upside of MCMC fitting and Bayesian analysis.
Fitting the same oscillation signals using both the nonlinear function and an exponential function yields similar oscillation periods within error margins. This suggests that the previously measured periods in the 2019 Catalogue remain reliable and applicable to MHD seismology. 
The relative plausibility between models is quantified by the Bayes factor, which compares Bayesian evidence (marginal likelihood). For synthetic data with varying noise levels generated by the nonlinear function, the Bayes factor consistently gives very strong evidence that favours the nonlinear model over the exponential one, demonstrating the detectability of nonlinearity. More importantly, the decaying kink oscillation observed on 2017-09-07 is better fitted by the nonlinear model than by either a Gaussian-exponential profile or an exponential damping profile. This indicates that the oscillation was nonlinearly damped by turbulence.
A potential concern is that the nonlinear function, having too many free parameters, may risk overfitting. However, the adopted Bayesian approach prevents this issue as the marginal likelihood balances the fit against model complexity \citep{Farrell_Lewandowsky_2018}. In the 2017-09-07 event, the marginal likelihood for the 6-parameter nonlinear model exceeds that of the 5-parameter exponential damping model, justifying the extra complexity for the observed data. %is warranted

We estimate the damping time of nonlinear kink oscillation during the first oscillation cycle. Our nonlinear turbulence damping model predicts \textbf{an} inverse proportional relationship between the quality factor and amplitude.
Using the nonlinear model to explain the empirical dependence of quality factor on $V_i/R$, compared with the resonant absorption model by Bayesian comparison, the regimes of nonlinearity and linearity are identified. Nonlinear damping is more likely to produce large quality factors at small $V_i/R$, low $\tau/P$ at high $V_i/R$, and very low $\tau/P$ regardless of amplitude, whereas resonant absorption is more plausible for damping times a few times the period in the small amplitude regime. 
This is an overall prediction based on an average over the full parameter space. 
%\sout{If discrepancies arise between prediction and observations, potential reasons include} \sout{ (1) non-uniform prior distributions of parameters; (2) the likelihood function that deviates from Gaussianity,} \sout{ or (3) incorrect assumptions about the default parameter range of parameters.} \textcolor{magenta}{I would remove all of this. There will always be discrepancies between predictions/observations because all our models are wrong. The three reasons you state are basically the same reason, prior assumptions. The likelihood function is also a prior, a prior on how we believe data were created. This is an idea that makes many researchers feel uneasy, but that's it. The advantage of the method is that all assumptions are explicit and hence their influence can eventually be evaluated.}
Applying the same model comparison technique to two distinct events with fixed observables, including density contrast and period, the results point to the nonlinear nature of the 2017-09-07 event and the linear nature of the 2012-05-30 event, which agrees with the time-series fitting outcomes.
It is important to note that the damping time used in our nonlinear model applies only to the first cycle. Nonlinear damping by turbulence involves four characteristic time scales, and identifying which one dominates in the middle or later phases is beyond the scope of this study.

Our next step is to review the events in the 2019 Catalogue to assess nonlinearity using MCMC fitting of time-series data and a Bayesian comparison of the damping quality factor. These two approaches will be applied in parallel to ensure consistency in the findings, and statistical conclusions will be extracted.
Once the nonlinear nature of a kink oscillation is established, seismological inversion based on the nonlinear theory can be applied to infer the local condition of loop plasma such as the density contrast and, consequently, the absolute magnetic field strength, by assigning a typical value to $\rho_e$, say $\rho_e=0.1\times 10^{-15}$~kg~m$^{-3}$. %Note that here density contrast obtained by time-series fitting may lack accuracy; however, it can be further constrained by the ratio of $P_{\rm k}$ to $P_{\rm A}$, which can be accurately determined through the fitting. 
The value of density contrast is the additional information that can be obtained from the nonlinear model, compared to the resonant absorption damping theory. With the statistical survey, it is worth investigating whether the value of $\eta$ differs with the choice of spectral line. In the 2017-09-07 event, the nonlinear fit yields $\eta = 0.2$, indicating that the detected oscillating plasma is confined within regions exceeding 80\% of the internal loop density. This aligns well with expectations for the 171\,\AA\ channel, which is more sensitive to the loop core \citep{2016ApJ...830L..22A}. This may imply that in the case of a multi-thermal loop, the observed damping and corresponding $\eta$ values may differ across spectral channels. 
In the current model, we consider the motion of the loop’s centre-of-mass, whereas in observations, the centre of emissivity is measured. In simulations, while the oscillation of the centre-of-mass and the centre-of-emissivity show minor differences in amplitude and period, they exhibit a noticeable divergence in frequency drift during the later stage. This effect will be investigated in detail in a follow-up study.

Is there a parameter that serves as the proxy for nonlinearity? It depends on which specific nonlinear effect to look at. The conjecture is that the larger the velocity, the more likely the oscillation is to be nonlinear. The nonlinearity parameter given by \cite{2014SoPh..289.1999R} is $\xi_0/R$ (equivalent to $VL/R$ where $L$ is the loop length), and by \cite{2010PhPl...17h2108R} is $V/l$. The parameter value greater than 1 is considered to be strongly nonlinear. In our model, theoretically, the magnitude of displacement amplitude over the loop radius is positively correlated to $V$ and $P$, hence $V$ could be an indicator of nonlinearity. Observationally, in the two events analysed in this work, the one that favours nonlinearity has a greater value of $V_i$. %see fig_evidence.pdf;Taking $L=100$~Mm, $V_i/R$ greater than 0.01 are considered to be nonlinear. 
%\textcolor{magenta}{comment: In the two events analysed in detail here, the one for which nonlinearity is more plausible has larger values of V$_i$, C$_1$, and density contrast.}
The criterion of KHI involves velocity amplitude and density contrast \citep{2019ApJ...870..108B,2019MNRAS.482.1143H}, and KHI could develop in loops with small oscillation amplitudes \citep[e.g.][]{2014ApJ...787L..22A}.

%Existing theories use amplitude as an input to predict the damping time (Tom). In our model, the amplitude is an output, with the magnitude determined by $\xi/R\propto \log{(V_i/R)}$.

%In this work, the other damping mechanisms are excluded by setting a sharp transition in the straight loop as we only care about the role of KHI-induced turbulence. 
%Do multiple damping mechanisms such as mode coupling work simultaneously? 
%True. we do see the loop cross-section is deformed even when the loop is at the equilibrium \citep{2024ApJ...966...68H}, indicating that other modes ($m>2)$ are excited and coupled with the kink mode.  In the condition of sharp density profile, resonant absorption is prevented initially by the condition of sharp density profile and in later times it is suppressed by high convection term (see detailed explanation in \citealt{2024ApJ...966...68H}).
%In the case of an oscillating loop with a smooth profile, there is a critical point of driving amplitude where KHI take over the role from the resonant absorption \citep{2016A&A...595A..81M}.
%Whether other mechanisms such as phase mixing and thermal misbalance are involved is unknown? (Antolin et al. 2017?) All these can be investigated by setting a smooth profile.
%Terradas+ 2006: coupling of the fast waves and shear Alfven waves->resonance-->damping.

Our nonlinear damping model is an approximation and not a perfect representation of the physical process. How can it be improved? One approach is to refine the density profile within the mixing turbulent layer by increasing the polynomial order \citep{2019ApJ...885..101H} and imposing additional boundary conditions. 
Currently, our mixing theory assumes that the velocity shear magnitude $\Delta V$ remains constant. However, simulations reveal that $\Delta V$ is time-dependent. A time-varying shear magnitude could alter the growth rate of the mixing layer, affecting the damping efficiency of loop oscillations. As a result, our model may overestimate the damping time. Incorporating this time dependence into an analytical framework poses a significant challenge. % or justify the error between time-averaged and time-varing chain is minor.
In addition, the contribution of mode coupling to damping requires further investigation \citep[e.g.][]{2021ApJ...910...58V}. In simulations, we spot that the loop cross-section is deformed even when the loop is at the equilibrium \citep{2024ApJ...966...68H}, suggesting that other modes ($m>2)$ are excited and coupled with the kink mode. 
Another direction is to make the loop model more realistic by accounting for thermal evolution and geometric effects such as loop curvature \citep{2004A&A...424.1065V,2024A&A...687A..30G}, loop expansion \citep{2008A&A...486.1015V}, and density stratification \citep{2005A&A...430.1109A}. However, whether these factors significantly influence KHI dynamics must first be determined. Notably, numerical simulations have shown that magnetic twist can suppress KHI \citep{2009ApJ...694..502O,2015ApJ...813..123Z,2018ApJ...853...35T}.
All these effects can be explored using high-resolution 3D MHD simulations on loop oscillations. %generalisation

In conclusion, we present an analytical description of how KHI-induced turbulence damps kink oscillation in coronal loops. Our analytic formula is validated by fitting the model to observational data supplemented by the Bayesian approach. We propose macroscopic features that are detectable within the current resolution limits and provide tools to identify nonlinear oscillation events. %\textcolor{red}{[Additionally, we provide a foundation for the seismological diagnostics based on this theory.]}
An alternative validation method to the theory is the forward modelling of simulations, which is currently in preparation as a follow-up study.

\begin{acknowledgments}
\textbf{Acknowledgments}\\
This work is supported by STFC Research grant No.~ST/Y00230X/1 and by project PID2024-156538NB-I00 from the Spanish Ministerio de Ciencia, Innovaci\'on y Universidades and FEDER funds. 
For the purpose of open access, the author has applied a Creative Commons Attribution (CC BY) licence to any author accepted manuscript version arising.\\
\textbf{Data Availability}\\
\noindent The synthetic data and fitting algorithm are available at \url{10.5281/zenodo.15639588} or \url{https://github.com/Sihui-Zhong/nonlinear-kink-oscillations/}. Observational data from AIA can be accessed via JSOC (\url{https://jsoc.stanford.edu/}). The active branch of the (P\underline{I}P) code is available at \url{https://github.com/AstroSnow/PIP}. Simulation data are available upon reasonable request.
\end{acknowledgments}

\vspace{5mm}
\facilities{SDO/AIA}

\software{SciPy \citep{2020NatMe..17..261V}, 
          SSWIDL \citep{1998SoPh..182..497F}
          %PIPpy (\url{https://github.com/AstroSnow/PIPpy})
          }



\appendix
\section{Sine and Cosine integral function}\label{Appendix:si(x)}
\begin{figure}[ht]
	\centering
	\includegraphics[width=\linewidth]{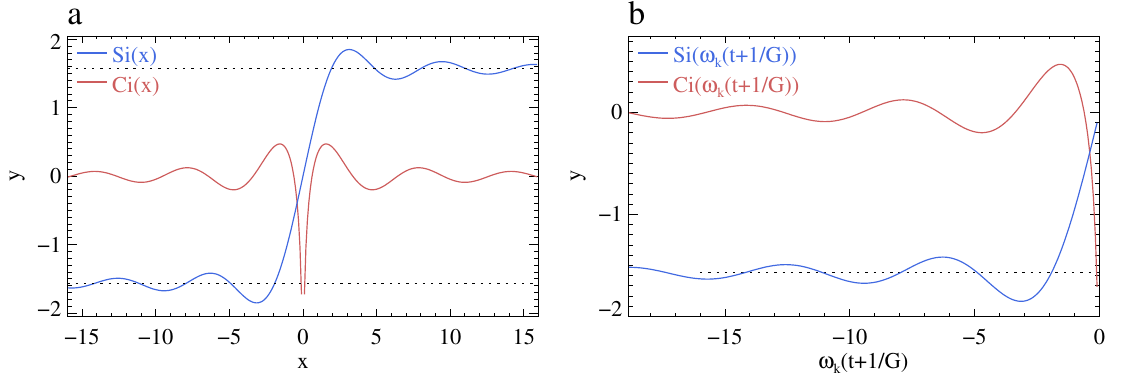}
	\caption{
        The sine (blue) and cosine integrals (red) as a function of $x$ (a) and $\omega_{\rm k}(t+\frac{1}{G})$ (b) as taken in Eq.~\ref{eq:nl_amplitude}. Here $\omega_{\rm k}=5$~mHz, $G=900$~s, calculated from a loop with $\rho_i/\rho_e=3$.
	}
	\label{fig:si(x)}
\end{figure}

A graph of sine and cosine integrals, Si(x) and Ci(x), is shown in Fig.~\ref{fig:si(x)}. For Python users, the function can be found in \texttt{scipy.special.sici}. 

\section{Parameter relation}
\label{sec:reduction}
The function of the oscillation amplitude (Eq.~\ref{eq:nl_amplitude}) contains 10 parameters, but not all of them are independent. This fact can be used to reduce the number of parameters in the function. In particular, it was estimated by \citet{2019ApJ...885..101H, 2024ApJ...966...68H} that $\Delta V \approx \frac{4}{3}V_i$. 
Here, we investigate their dependence by running a set of low-resolution simulations of standing transverse oscillations of a magnetic tube driven by various initial velocities $V_0$ as follows. 

Following \citet{2024ApJ...966...68H}, we model half of a straight magnetic loop (from footpoint to apex) with radius $R$ and internal density $\rho_i$ surrounded by a medium with $\rho_e$ in Cartesian coordinates. The loop is impulsively driven by a velocity perturbation in the $x$-direction with $V_0$ ranging from 0.05 to 0.3 (cf, the sound speed of the corona $C_s=1$) to kick off the transverse oscillations. The effect of density contrast inside and outside the loop is also considered by setting $\rho_i/\rho_e=2,3,5,10$, which are typical values in the coronal environment. 
The simulation is performed in the ideal MHD scheme with 3 dimensions using the (P\underline{I}P) code \citep{2016A&A...591A.112H}. The set of MHD equations is the same as that in Section 2.1 of \citealt{2024ApJ...966...68H}. The setup, including initial velocity perturbation, density distribution, and boundary conditions, is also the same as that in \citet{2024ApJ...966...68H}. The loop footpoint is tied to obtain a standing oscillation. The tube has a homogenous inner core and a sharp transition to the external region, which is given by the hyperbolic tangent function. We set the thickness to be at the limit of the resolution preventing resonant absorption in the numerical simulation (on top of resonant absorption being a slow process in very thin layers). We take $R=0.5$, $\rho_e=1$, $\rho_i=[2,3,5,10]$, $V_0\in[0.05,0.3]$, $\beta=0.05$, $T_e/T_i=2$. The magnetic field is uniform and parallel to the loop axis ($z$). Since we only care about the velocity profile and do not focus on the KHI vortices which have to be captured by high-resolution grids, our simulation uses a low grid resolution with $144\times128\times52$ grid points in 3 dimensions. The length scale in the 3 domains is $x=[-L_x,L_x],y=[0,L_y],z=[0,L_z]$ where $L_x=2.7, L_y=2.4, L_z=30$.

\begin{figure}[ht]
	\centering
	\includegraphics[width=0.6\linewidth]{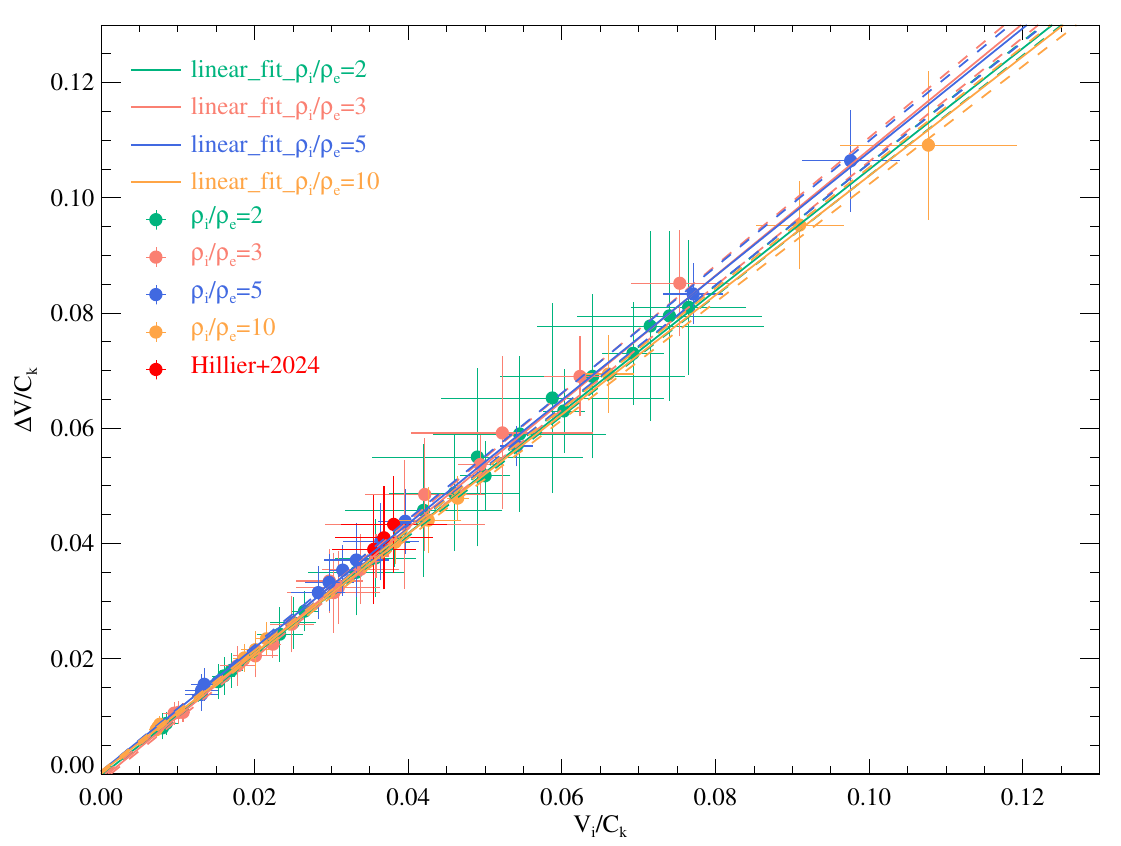}
	\caption{
        The dependence of shear magnitude ($\Delta V$) on velocity ($V_i$) of the oscillating loop as measured in a set of simulations. Each colour represents a case of $\rho_i/\rho_e$, and the x/y error is indicated by the length of the horizontal/vertical bar across the data symbol. Data for each case are fitted with a coloured solid line with resulting slopes 1.05, 1.09, 1.08, 1.04 for $\rho_i/\rho_e=2,3,5,10$ respectively. The dashed lines show the slope error. $C_k$ is the kink speed calculated under the long-wavelength limit.
	}
	\label{fig:dV-scaling}
\end{figure}

In total, we performed 27 runs. For each case, we measure the representative $V_i$ and $\Delta V$. In mixing theory, $\Delta V$ is assumed to be constant in time. However, as observed in simulations, $\Delta V$ varies as the oscillation evolves, so we take a time-averaged value as representative for the mixing theory. 
The measurement of $\Delta V$ is done by first extracting the velocity profile across the loop at the $y$-axis at 2 or 3 different times when the momentum is around zero (i.e., the loop is at the equilibrium position), then calculating $\Delta V$ as the velocity difference inside ($V_i$) and outside the loop at each instant of time and finally taking the average as representative. 

Fig.~\ref{fig:dV-scaling} shows the scaling of $\Delta V$ with $V_i$, which is linear. Data are plotted with error bars and fitted with a linear function using the orthogonal distance regression in Python \texttt{(Scipy.odr)} in each case of $\rho_i/\rho_e$. All the best-fitted lines virtually cross the zero origin. All intercepts are smaller than 0.001, hence approaching zero. For $\rho_i/\rho_e=2,3,5,10$, the gradient of the best-fitted line is 1.05, 1.09, 1.08, and 1.04, respectively. The errors in gradients, depicted by the dashed lines, are around 0.01 on average. The linear fit of all data results in a slope of $1.063\pm0.007$. In general, $\Delta V$ is linearly dependent on $V_i$ with a factor of around 1.1, hence $\Delta V = 1.1V_i$. We measured the $V_i$ and $\Delta V$ data-pairs in the simulation of \citet{2024ApJ...966...68H} and included them in the same plot, see the red dots, and they are consistent with the obtained scaling. %\textcolor{red}{Hence, from now on the model in Eq.~\ref{eq:nl_amplitude} will consist of 9 parameters.}

%The scaling of $\Delta V$ with $V_i$ is obtained through a set of low-resolution simulations of transverse oscillations of a magnetic tube with $\rho_i$ driven by various $V_i$, see Fig.~\ref{fig:dV-scaling}.

\section{Curve fitting with synthetic data}\label{appendix:curvefit}
As a preliminary step, curve fitting by the model time series is first tested on synthetic data before being applied to real observations. The tests aim to see if the fitting procedure can recover the input parameters and distinguish the nonlinear and linear models if the signals are contaminated with noise.
We create synthetic oscillation signals with our nonlinear damping model (Eq.~\ref{eq:nl_fitting}), contaminated with different levels of noise, and fit the signals with Eq.~\ref{eq:nl_fitting}-\ref{eq:exp} by the MCMC method. Noise is generated by a series of random numbers drawn from a normal distribution with zero mean and a standard deviation between $0-30\%$ of the oscillation amplitude, generated using \texttt{randomn.pro}. 

\begin{figure}[ht]
	\centering
	\includegraphics[width=\linewidth]{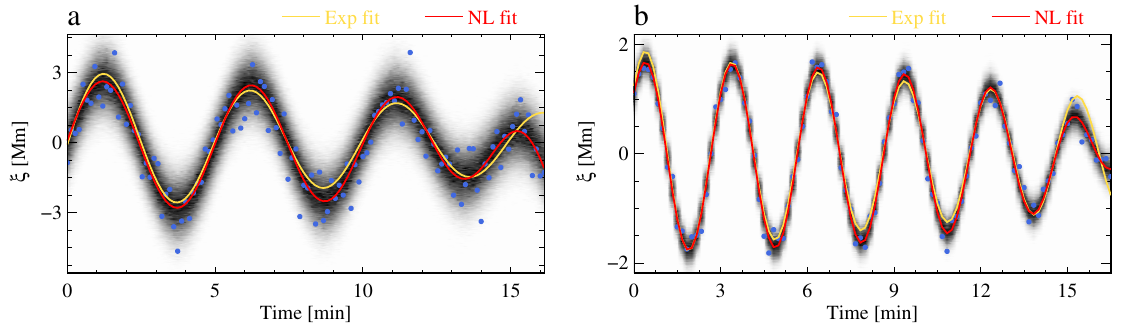}
	\caption{
        The best fits to the synthetic signals (blue dots) by our nonlinear damping function Eq.~\ref{eq:nl_fitting} (NL, marked by red curve) and exponential damping function (Exp, marked by yellow curve). The signals are produced using Eq.~\ref{eq:nl_fitting} with $t_0=0$~s, $R=2$ Mm, $\rho_i/\rho_e=3$, $C_1=0.3$, $V_i=60$ km\,s$^{-1}$, $P_{\rm k}=300$ s, $P_{\rm A}=268$ s, $\Delta m=-0.85$, $\Xi=0.24$ and noise level of 30\% for the 4-cycle signals (panels a) and with $t_0=-20$ s, $R=2.5$ Mm, $\rho_i/\rho_e=3$, $C_1=0.3$,$V_i=60$ km\,s$^{-1}$, , $P_{\rm k}=180$ s, $P_{\rm A}=161$ s, $\Delta m=-0.85$, $\Xi=0.24$ and noise level of 10\% for the 6-cycle signals (panels b). The meaning of the shading is the same as that in Fig.~\ref{fig:fit_obs}. An example of the selected ranges for priors in MCMC fitting are as follows: $t_0\in[-10,10]$~s, $\rho_i/\rho_e\in[1,10]$, $C_1/R\in[0.01,0.3]$~Mm$^{-1}$, $V_i\in[30,100]$~km~s$^{-1}$, $P_{\rm k}\in[100,500]$~s, $\eta\in[0,1]$. 
        }
	\label{fig:fit_synthetic}
\end{figure}

Here, two series of signals are generated: a 4-cycle one with $t_0=0$, $R=2$ [Mm], $\rho_i/\rho_e=3$, $C_1=0.3$, $V_i=60$  [km\,s$^{-1}$], $P_{\rm k}=300$ [s], $P_{\rm A}=268$ [s], $\Delta m=-0.85$, $\Xi=0.24$ and a longer one with 6 cycles with $t_0=-20$ [s], $R=2.5$ [Mm], $\rho_i/\rho_e=3$, $C_1=0.3$,
$V_i=60$ [km\,s$^{-1}$],  $P_{\rm k}=180$ [s], $P_{\rm A}=161$ [s], $\Delta m=-0.85$, $\Xi=0.24$.
An example for each time series and the corresponding fitting results with two functions are shown in Fig.~\ref{fig:fit_synthetic}. Looking at panels a and b, the best-fitted nonlinear curves (red) match the signals pretty well. The posterior predictive distribution for the nonlinear model (in grayscale) covers the signals. 
This demonstrates that MCMC fitting with a function with 7 parameters is achievable. The best-fitted parameter values with 95\% credible intervals are displayed in Table~\ref{table:nl_pars}. 
For clean signals (zero-noise), the values of the parameters are almost recovered in the best fit. For signals with a noise level increasing to 30\%, the measurement of kink periods and driving velocity is accurate, with a Gaussian-like narrow-band distribution. 
However, other parameters are not constrained that well, as it is their combination that controls the oscillatory pattern, rather than a single variable. When fitting with an observed oscillation, a more informative prior distribution can be set for observables such as $R$ and $V_i$ by rough measurements in observations.

\begin{table}[ht]
\centering
\begin{tabular}{lcccccccc} 
 \hline
 Noise & $C1/R$ [Mm$^{-1}$] & $\rho_i/\rho_e$  & $V_i$ [km\,s$^{-1}$] & $P_{\rm k}$ [s] &  $\eta$ & $t_0$ [s] & $P_{\rm e}$ [s] & $BF_{\rm NL/Exp}$ \\  
 \hline\hline
 S1 & 0.15 & 3 & 0.3 & 60 & 300 & 0.14 & 0 \\
 \hline
 0      & $0.15^{+0}_{-0}$ & $3.0^{+0}_{-0}$  & $60^{+0}_{-0}$ & $300^{+0}_{-0}$   & $0.14^{+0}_{-0}$  & $0^{+0}_{-0}$ & $295^{+3}_{-3}$ & 193 \\ 
 10\%   & $0.14^{+0.02}_{-0.01}$ & $1.9^{+2.1}_{-0.2}$ & $62^{+2}_{-4}$ & $296^{+3}_{-2}$  & $0.17^{+0.04}_{-0.06}$ &   $3^{+2}_{-3}$  & $294^{+3}_{-3}$ & 116 \\
 20\%  & $0.14^{+0.01}_{-0.13}$ & $2.4^{+1.5}_{-0.5}$ & $66^{+6}_{-4}$ & $300^{+5}_{-4}$  & $0.18^{+0.13}_{-0.05}$ &  $0^{+5}_{-5}$ & $294^{+4}_{-4}$ & 49 \\
 30\%   & $0.15^{+0.02}_{-0.02}$ & $2.6^{+1.5}_{-0.8}$ & $57^{+8}_{-6}$ & $303^{+6}_{-8}$ & $0.13^{+0.15}_{-0.06}$ &   $-4^{+10}_{-7}$ & $297^{+7}_{-4}$ & 32 \\
 \hline\hline
 S2 & 0.12 & 3 & 0.3 & 60 & 180 & 0.14 & $-20$ \\ 
 \hline
 0    & $0.12^{+0}_{-0}$ & $3.0^{+0.0}_{-0.0}$  & $60^{+0}_{-0}$ & $180^{+0}_{-0}$  & $0.14^{+0}_{-0}$  & $-20^{+0}_{-0}$ & $179^{+1}_{-1}$ & 1050 \\
 10\%   & $0.13^{+0.01}_{-0.02}$ & $3.0^{+0.4}_{-1.8}$ & $60^{+4}_{-2}$& $180^{+1}_{-1}$  & $0.09^{+0.10}_{-0.02}$  & $-21^{+2}_{-2}$ &  $179^{+1}_{-1}$ & 54 \\
 20\%   & $0.12^{+0.02}_{-0.08}$ & $3.9^{+0.8}_{-1.63}$ & $64^{+7}_{-6}$ & $178^{+2}_{-1}$ & $-0.17^{+0.79}_{-0.07}$  &  $-17^{+3}_{-4}$ & $178^{+1}_{-1}$ & 9 \\
 30\%   & $0.13^{+0.07}_{-0.04}$ & $2.5^{+2.3}_{-1.2}$ & $54^{+15}_{-6}$ & $184^{+5}_{-4}$  & $0.26^{+0.72}_{-0.13}$  & $-28^{+8}_{-7}$ & $181^{+2}_{-2}$ & 7  \\
 %\hline\hline
%\multicolumn{9}{l}{2017-09-07 event}\\
%\hline
%  & $0.04^{+0.01}_{-0.01}$ & $5.2^{+0.5}_{-0.6}$ & $128^{+8}_{-9}$ & $482^{+5}_{-4}$  & $0.21^{+0.07}_{-0.05}$ & $-1^{+7}_{-7}$  \\
%\hline
%  \multicolumn{9}{l}{2012-05-30 event}\\
%\hline
%  & $0.04^{+0.01}_{-0.01}$ & $3.1^{+0.2}_{-0.1}$ & $93^{+8}_{-7}$ & $254^{+3}_{-3}$ & $0.99^{+0.01}_{-0.1}$  & $-7^{+5}_{-5}$  \\
%   & $0.05^{+0.01}_{-0.01}$ & $3.2^{+0.2}_{-0.1}$ & $88^{+6}_{-11}$ & $254^{+4}_{-4}$ & $0.99^{+0.01}_{-0.18}$  & $-9^{+8}_{-8}$  \\
 \hline
\end{tabular}
\caption{The best-fitted parameters obtained by MCMC fitting with Eq.~\ref{eq:nl_fitting} to the synthetic signals, and Bayesian model comparison between nonlinear turbulence damping model ($\mathbf{M}_{\rm NL}$, Eq.~\ref{eq:nl_fitting}) and exponential model ($\mathbf{M}_{\rm Exp}$, Eq.~\ref{eq:exp}). S1 and S2 are the two series (corresponding to panel a and b of Fig.~\ref{fig:fit_synthetic}, respectively) with inputs shown in the same row. Rows below the inputs are the best-fitted values to the time series for different levels of noise. The noise level is estimated as the ratio of the noise standard deviation of the maximum amplitude of the clean oscillatory signal. $P_e$ is the best-fitted oscillation period for $\mathbf{M}_{\rm Exp}$. $BF_{\rm NL/Exp}$ is the Bayes factor. A Bayes factor greater than 2/6/10 indicates a positive/strong/very strong preference towards the nonlinear model. The errors are determined by the 95\% credible intervals.}
\label{table:nl_pars}
\end{table}

The best fits of the exponential damping function to the two time series are shown by the yellow curves in Fig.~\ref{fig:fit_synthetic}.  
Quantitative assessment of the relative plausibility of two competing models is computed using the Bayes factor, a measure of the strength of the evidence in favour of one model over another. 
As displayed in Table~\ref{table:nl_pars}, all results give strong to very strong evidence towards our nonlinear model. This indicates that Bayesian inference can distinguish such a complicated nonlinear model from the linear model. 
In addition, the oscillation periods measured by both fitting functions are close enough. This indicates that the linear damping model can well capture the period of nonlinear oscillations, since the kink period is dominated in the initial phase of the oscillation, and period drift most clearly manifests in the last cycle. Hence, the oscillation periods in the 2019 Catalogue that are measured using the linear model are reasonable.

\begin{figure}[ht]
	\centering
    \includegraphics[width=\linewidth]{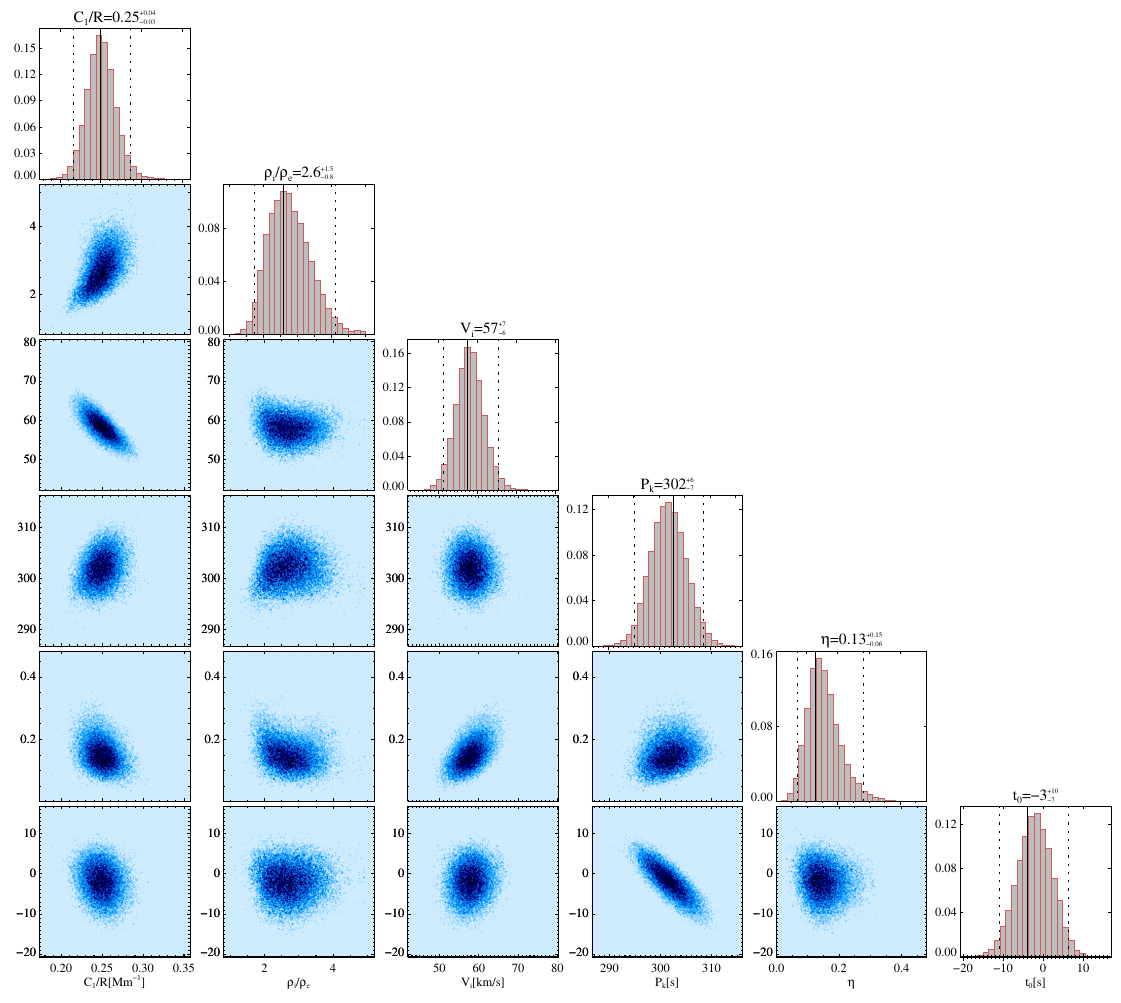}
	\caption{
       Corner plot showing the marginal posterior distributions for the parameters in the nonlinear function (Eq.~\ref{eq:nl_fitting}) along the diagonal and the joint posterior distributions showing correlations between two certain parameters under the diagonal. Data are obtained via MCMC sampling during the fitting procedure for 4-cycle synthetic oscillation with 30\% noise (Fig.~\ref{fig:fit_synthetic}a). The histograms show the posterior density probability of each parameter, with the black vertical lines indicating the best-fitted values. The 2D density plots show the covariances between variables. An elongated shape suggests a strong correlation, while a circular or rectangular shape implies no correlation. 
	}
	\label{fig:correlation}
\end{figure}

%[Parameter correlation]\\
One advantage of Bayesian analysis, supplemented with MCMC sampling, is that it allows us to explore the internal relationships between fitting parameters based on a large number of samples.
Fig.~\ref{fig:correlation} shows the distribution of each parameter and the 2D joint posterior distribution of samples of two selected parameters among 6 for the 4-cycle oscillations with 30\% noise. Such plots reveal correlations and degeneracies among the parameters as expected. This analysis is based on the $10^5$ samples generated during the MCMC fitting. Looking at the histograms at the end of each row, most parameters are well constrained, with some showing strong dependencies.  
In particular, a negative correlation between $C_1/R$ and $V_i$ indicates that higher probable values of $C_1/R$ correspond to smaller possible values of $V_i$. This is because $\Delta V$ and $C_1/R$ are tied together in the form of $C_1\Delta V/R$ in Eq.~\ref{eq:velocity_amp}.
%a positive correlation between $R$ and $C_1$ indicates that higher probable values of $R$ correspond to larger possible values of $C_1$. This also implies that taking a different prior distribution for $C_1$ can result in different best-fit values for $C_1$ and $R$. This is borne out by running the fits with a uniform prior distribution for $C_1$, leading to an increase in the preferred values obtained by the fitting routine for $C_1$ and $R$. However, the preferred values for the remaining parameters were not significantly affected by this change in prior. 
Additionally, $P_{\rm k}$ is anti-correlated to $t_0$ as variations in $P_{\rm k}$ can be offset by shifts in $t_0$ to preserve the phase alignment between the model and the observed oscillation.

\bibliography{references}{}
\bibliographystyle{aasjournal}



\end{document}